\documentclass[manuscript]{aastex}

\slugcomment{Not to appear in Nonlearned J., 45.}

\shorttitle{}
\shortauthors{Lin et al.}

\begin{document}

\title{An investigation of CME dynamics, productivity and 
source-region subsurface structure}

\author{C.-H. Lin}
\affil{Graduate Institute of Space Sciences, National Central University,
    Chung-Li, Taoyuan, Taiwan}

\email{chlin@jupiter.ss.ncu.edu.tw}

\author{C. S. Baldner}
\affil{Hansen Experimental Physics Laboratory, Stanford University,
Stanford, CA94305-4085, U.S.A.}

\author{J. Chen}
\affil{Naval Research Laboratory, Plasma Physics Division, Washington, DC20375, U.S.A.}



\begin{abstract}

This work is to
study the dynamics of coronal mass ejections and
to investigate possible
correlation between CME productivity and 
subsurface structure.
Two CMEs and six active regions are selected for the study.
The CMEs are examined by comparing observations and theoretical models,
and the subsurface structures are probed by local helioseismic inversions.
The analysis of the CMEs shows that
the eruptive flux-rope model is in good agreement with both events.
However,
some discrepancies with the observation are also found,
indicating that the model can be further improved.
The helioseismic investigation results
indicate a consistent correlation
between the CME productivity and the subsurface temperature structure.
The inferred subsurface magnetic structure reveals that
the source regions of the two studied CMEs may share similar 
subsurface structures.
Not found in the other CME productive regions selected for consideration,
however, the similarity in the subsurface features may indicate
the presence of a subsurface structural connection between
the two closely located and CME-productive regions.
\end{abstract}


\section{Introduction}
The dynamics of the solar corona 
are ultimately governed by the dynamics of the solar magnetic field,
which is generated in the interior and
brought up across the photosphere into the corona
by plasma motion.
As a result,
the subsurface structure and processes are likely to influence
the dynamics in the corona.
Among all observed solar dynamical activities,
coronal mass ejections (CMEs),
which eject large amounts of mass and magnetic flux
from the Sun to interplanetary space, 
are the most spectacular events.
They release energy in the order of $10^{32}-10^{33}$~erg 
during the process.
How such violent eruptions are initiated and driven,
and how the processes are related to the subsurface properties 
of their source regions
are, however, still under debate.

Several theories have proposed that 
CMEs are fast expanding magnetic flux ropes
driven by
the twisting of the flux-rope footpoints
caused by certain processes below the photosphere.
The advances in helioseismology and observation
have enabled examinations of local subsurface structures and dynamics
at and around a small target region.
Several groups have utilized the tool
to search for the subsurface signatures of such twisting,
which would facilitate better forecasting and understanding of
large flares and CMEs
\citep{Komm_etal_2004ApJ,KH2009JGRA,Komm_etal_2011SoPh,Reinard_etal2010ApJ,Mason_etal2006ApJ}.
The common strategy has been to look for
a fluid dynamics parameter
that can differentiate active regions with different flaring activity levels.
For instance,
\citet{Komm_etal_2011SoPh} found that
the structure component of vorticity,
when combined with the total surface flux,
can be used to predict C- and M-class flares.
\citet{Reinard_etal2010ApJ}, using another parameter,
Normalized Helicity Gradient Variance (NHGV),
reported that NHGV is able to separate non-flaring from flaring ARs
and also to forecast and distinguish C-, M- and X-class flares.
Since large flares are often associated with CMEs,
\citet{Webb_etal2011SoPh} suggested that such subsurface parameters
may be related to CME driving mechanisms.

In contrast to the progress on the connection between
flares/CMEs and the subsurface flows of active region,
little work has been done to examine 
the correlation between 
the productivity level of flares and CMEs (or eruptivity, in short)
and
the subsurface structural properties
that are not dynamical 
on the timescale of coronal eruption,
which is what we wish to explore in this paper.
Due to the large differences in physical parameters below and above 
the photosphere,
the characteristic 
time scales are expected
to be longer in the convection zone than in the corona.
Thus, 
the non-dynamical part of the subsurface structure is 
unlikely to show detectable change over
the time scale of one eruption in the corona.
However, different eruptivity levels in the corona may be a result
of different subsurface structures of the source regions.

In this study, 
our objectives are two fold:
on the one hand,
we aim to better understand how CMEs are initiated and driven
by testing various proposed driving mechanisms using
observed CME data;
and on the other hand,
we explore the possibility of association, if any, 
between the CME productivity
and the magnetic/thermal structures below their source regions. 
Two prominent CMEs and six active regions (ARs)
are studied in this work.
The dynamics and possible eruption mechanisms of the CMEs
are examined
by comparing observations with three different models.
The subsurface thermal structures of the six ARs, 
which include the source regions of the chosen CMEs, 
are probed by
the local helioseismic method 
\citep{Hill1988ApJ,Basuetal2004ApJ}.
To infer the subsurface magnetic structure,
we use the ratio of magnetic pressure over gas pressure 
(i.e., $\beta_H \equiv P_{\rm mag}/P_{\rm gas}$)
to represent the significance of 
the magnetic-field effect relative to the gaseous structure.
Note that $\beta_H$ differs from the usual plasma $\beta$,
which is defined as $P_{\rm gas}/P_{\rm mag}$.
The result of the inferred subsurface thermal and magnetic structures
are depth profiles averaged over a 
$15^\circ$ patch in space and approximately 4 to 6.5 days in time.

The rest of the paper is organized as follows:
The observations and data are described in Sec.~\ref{sec:data},
the three selected CME models are reviewed in Sec.~\ref{sec:cmemdl},
the studies of the CMEs and of the subsurface structures are 
presented in Sec.~\ref{sec:cme} and \ref{sec:sub},
respectively,
and 
a summary of our results is given in Sec.~\ref{sec:sum}.

\section{Observation and data} \label{sec:data}
The selected CMEs are the most prominent two events
during the observation period
2008 March 25 -- April 5.
One was from AR10989 on March 25,
and the other was from AR10988 on April 5.
For the rest of the paper, 
we distinguish the CMEs by the month they erupted, 
with ``03'' and ``04'' referring to March and April, respectively;
that is, CME03 for the one on March 25 and CME04 for the one on April 5.
We utilized the image data from STEREO/SECCHI
(Solar TErrestrial RElations Observatory/Sun 
Earth Connection Coronal and Heliospheric Investigation; 
\citet{STEREO2008SSRv}) 
instruments (EUVI~171\AA~channel, COR1, and COR2)
to derive the kinematics.
We also incorporated the
soft X-ray (SXR) data from GOES 
(Geostationary Operational Environmental Satellite; 
\citet{GOES1994SoPh}) 1--8\AA~channel,
and hard X-ray (HXR) data from
RHESSI (Reuven RAmaty High Energy Solar Spectroscopic Imager; 
\citet{RHESSI2002SoPh})
for the associated flare information.

\begin{figure}
\includegraphics[width=14cm]{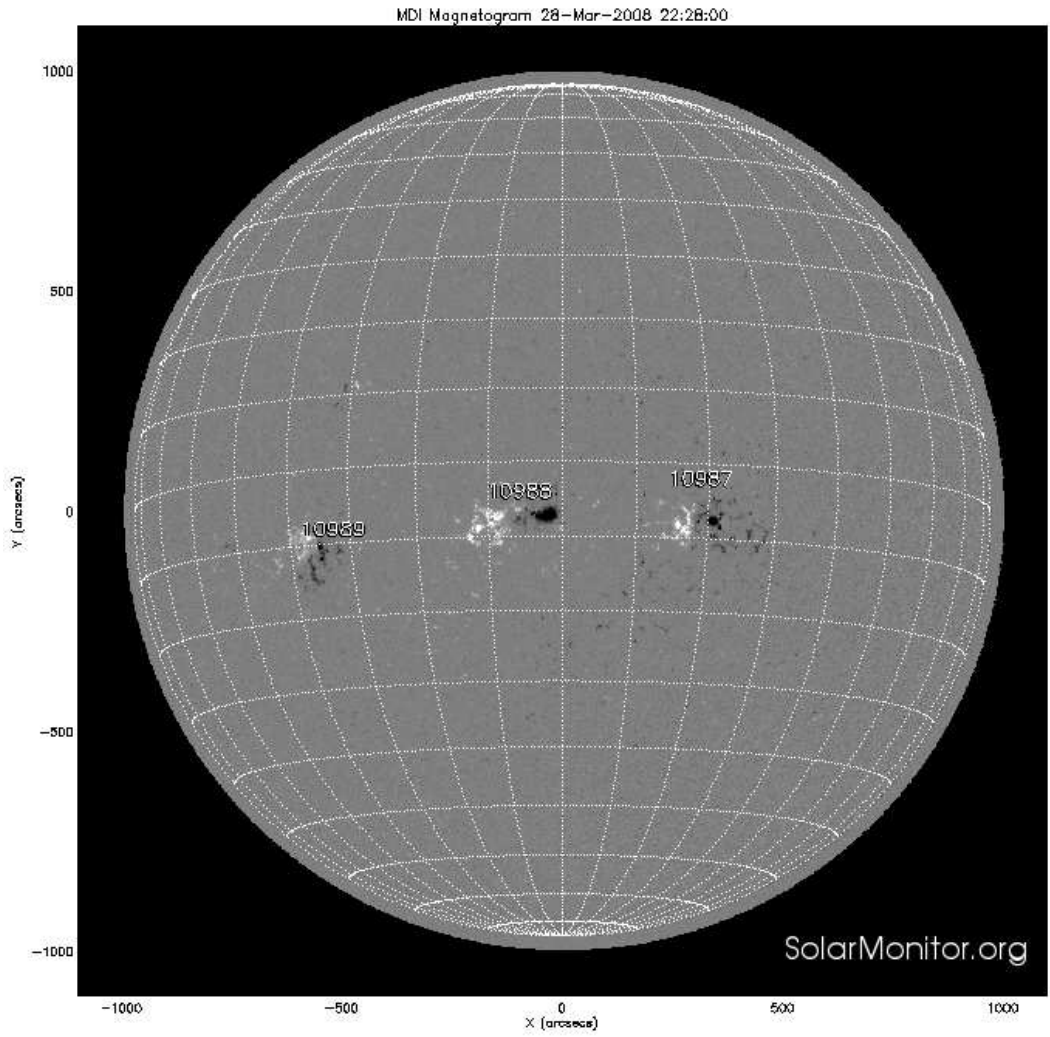}
\caption{A representative MDI magnetogram to illustrate 
the relative locations and surface magnetic fields of 
the three active regions.}
\label{fig:mdi}
\end{figure}

For the investigation of eruptivity and subsurface structure,
we examined the following six active regions:
AR10987--10989, AR09404, AR10980, and AR11123.
AR10987--10989
were the main ARs visible on the solar surface during 
the Carrington Rotation (CR) 2068.
The three ARs located at similar latitudes, were almost equally spaced
(cf. Fig.~\ref{fig:mdi}),
but showed different levels of eruptivity:
AR10989 was the source region of CME03
and an associated M-class flare,
AR10988 was the source region of CME04, and
AR10987 produced neither large CME nor strong flare.
AR09404 was a mature region in CR1974.
During its crossing, 2001 March 27 -- April 5,
there were more than 16 ARs on the solar surface,
and some of them were very eruptive.
However, AR09404 was relatively quiet
and produced only two modest flares.
AR10980 was a decaying AR, and produced
one CME and an associated C-class flare on 2008 Jan 2.
It had also produced several strong CMEs and flares in 
the previous rotation, CR2064, during which it was identified as AR10978.
AR11123 was a newly emerged AR that produced multiple CMEs and flares on
2010 Nov 11, 
which was during the tracking time period of our helioseismic inversion
(cf. Table~\ref{tab:tab_coord}). 
Our inclusion of ARs erupted within, before and after inversion intervals;
closely located ARs at similar life stage; and
ARs from different CRs at different stages of their lives
enable us to verify whether our results are
dependent on factors such as 
temporal sequence of eruption and inversion,
solar cycle, AR evolution and large scale structural differences.

Our helioseismology data consist of full-disk Dopplergrams from 
the Michelson Doppler Imager (MDI; \citet{MDI1995SoPh}) on board
the Solar and Heliospheric Observatory(SOHO) during SOHO/MDI dynamics runs, 
and the Dopplergrams from the Helioseismic and Magnetic Imager 
(HMI; \citet{HMI2012SoPh}) on board the Solar Dynamics Observatory (SDO).
We followed the same ring-diagram inversion procedure as described in
\citet{Basuetal2004ApJ}
to study the local structure below the active regions.
In short,
the ring-diagram analysis (Hill, 1988; Patron {\it et al.}, 1997)
was implemented
to determine the frequencies of short-wavelength (high degree)
modes in a small region of interest.
These are the modes that
can be approximated as plane waves over a small area of the Sun.
Hence, long-wavelength (low degree) modes are excluded,
and the inversion can only probe down to a certain depth.
In our case, the reliable inversion depth range
is approximately the upper 3\% of solar radius, 
or above $\sim 0.97R_\odot$.
The differences between the mode frequencies in an active region and 
in a reference quiet region 
were then input to the inversion procedure
to determine the difference in the structure
between the two regions.
For our inversion,
we cut a $15^{\circ}$ patch around each target AR
and also a patch around a nearby region of quiet Sun (QS) 
at the same latitude.
To reduce the errors due to the projection effects,
we tracked the target AR only while it is close to the disk center.
The exact tracking times and coordinates of the ARs
are listed in Table~\ref{tab:tab_coord}.
\section{CME models} \label{sec:cmemdl}
The models we selected for our study are
the catastrophe (CA) model 
\citep{PF1990SoPh,FI1991ApJ,FP1995ApJ,LF2000JGR,PFbook2000},
eruptive flux-rope (EF) model \citep{chen1989ApJ},
and breakout (BO) model \citep{antiochos1999ApJ}. 
An illustration of the three models is shown in Fig.~\ref{fig:CMEmdls}.
In the following, we give a review of the three models.

\subsection{Catastrophe model (CA)} \label{sec:mdl_CA}
The CA model 
is a two-dimensional flux-rope model derived under the condition of
ideal magnetohydrodynamics (MHD).
The configuration of the model,
which is obtained by solving the Grad-Shafranov equation,
consists of a flux rope in equilibrium and two line sources at the photosphere
(cf. Fig.~\ref{fig:CMEmdls}A).
The flux rope is driven out of balance when
the photospheric sources {\em slowly} move toward each other.
The motion at the photosphere is assumed to be much slower than
the Alv\'en time scale at the corona such that
the evolution of the flux rope can be considered as a sequence of
force-free equilibria \citep{PFbook2000}. 
The magnetic equilibrium of the system is lost
when the separation of the two sources becomes less than a critical distance.
The flux rope is then thrown upward by the imbalance between
the upward magnetic pressure and downward magnetic tension forces.
The height where this loss of balance 
happens is called
critical height, $\lambda_0$.
After the loss of equilibrium,
the behavior of the flux rope is determined by
the dynamics of the system.
In order for the erupting flux rope to escape the solar surface
to become a CME,
the model proposes that magnetic reconnections must occur;
otherwise, 
the flux rope would stop rising once 
it finds a new equilibrium and
its kinetic energy is dissipated.
To allow magnetic reconnections after
the initial eruption, which is an ideal MHD process,
the model proposes that a current sheet,
or X-line, forms below the erupting flux rope.
The field lines inside the current sheet
are no longer frozen to the plasma but can move through each other,
resulting in the reconnection.

Because the topology of the magnetic field configurations
changes after the formation of a current sheet
and after the magnetic connections,
the calculation of the model is divided into several stages 
\citep[see][for detail]{LF2000JGR}.
For the initial eruption when
the flux-rope radius can be considered to be much smaller than
the critical height,
\citet{PFbook2000} derived an analytical expression
for the velocity of the flux rope:
\begin{eqnarray}
\dot{h} \approx \sqrt{\frac{8}{\pi}}
\left[
\ln \left(\frac{h}{\lambda_0}\right) + \frac{\pi}{2} -
2 \tan^{-1}\left(\frac{h}{\lambda_0}\right)
\right]^{1/2} + \dot{h}_0,
\label{eqn:CA}
\end{eqnarray}
\noindent
where $h$ is the height of the flux rope,
$h_0$ is the initial height, and
$\lambda_0$ is the critical height.

In the model,
magnetic reconnection is assumed to occur
at the mid-point of the current sheet.
The magnetic reconnection rate is
taken to be the rate of magnetic flux crossing the reconnection point,
which
is related to the electric field in the current sheet
according to Faraday's law:
\begin{eqnarray}
E_z &=& -\frac{1}{c} \frac{\partial A(0,y_0)}{\partial t},
\label{eqn:recrate1}
\end{eqnarray}
where
$y_0$ is the reconnection point, 
$A(0,y_0)$ is the magnetic flux passing through the current sheet,
and $E_z$ is the electric field in the current sheet \citep[e.g.,][]{LF2000JGR}.
Therefore, $E_z$
is a direct representation of the prescribed reconnection rate.
In the model, $E_z$ is assumed to be the same everywhere 
along the current sheet.
$E_z$ can also be written
as $E_z = M_A \; V_A(0,y_0) \; B_y(0,y_0)/c$,
where $V_A$ is the Alfv\'en speed, 
and $B_y$ is the magnetic field along the current sheet.
$M_A \equiv |V_R/V_A|$ is Alfv\'en Mach number,
where $V_R$
is the velocity of the plasma flowing into the current sheet.
Hence, the reconnection rate
is essentially parameterized by $M_A$.
Different flux-rope motion can be produced by tuning $M_A$.
\citet{LF2000JGR} varied $M_A$ to examine 
the effect of the reconnection rate on the CME motion.
They found that
if $M_A$ is not greater than $0.001$,
the flux rope would oscillate around an equilibrium height
when the downward and upward forces become balanced again.
For an intermediate reconnection rate, that is, $0.001 < M_A < 0.041$,
the CME would go through a brief deceleration phase after
the initial loss of balance and then accelerate again
to reach peak acceleration.
When $M_A > 0.041$, the deceleration phase ceases to happen.

To identify a CME as driven by the CA mechanism,
we consider
if 
the initial eruption starts {\em before} the rise of SXR emission
and can be described by Eq.~\ref{eqn:CA};
the rest of the kinematics can qualitatively be re-produced
by tuning $M_A$ in the manner as described by \citet{LF2000JGR};
and
the signatures of the magnetic reconnections 
appear {\em below} the flux rope and
{\em after} the on-set of the eruption.
As an example for the last point,
\citet{PFbook2000} found in their simulation that
the reconnection rate
did not rise above zero until seven minutes after the initiation.

\subsection{Eruptive flux-rope model (EF)} \label{sec:mdl_EF}
The EF model is formulated based on 
ideal MHD
\citep{chen1989ApJ}.
The two dominant forces acting on a flux rope in the initial equilibrium are
an outward Lorentz self-force, 
resulting from the toroidal electric current ($I_t$) in the flux rope,
and a downward Lorentz force due to the 
flux-rope toroidal magnetic field $B_t$ and the
ambient coronal magnetic field.
An increase of $I_t$, causing an increase of the outward force,
would destablize the flux rope, leading to an accelerated expansion.
In the model,
the increase of $I_t$ is represented by 
the increase of
the poloidal magnetic flux $\Phi_p \equiv cL(t)I_t(t)$ of the flux rope,
where $c$ is the speed of light and $L$ is the self-inductance of
the current system.
In short,
the eruption results from 
a positive $d \Phi_p(t)/dt$.

\begin{figure}
\begin{center}
\includegraphics[height=20cm]{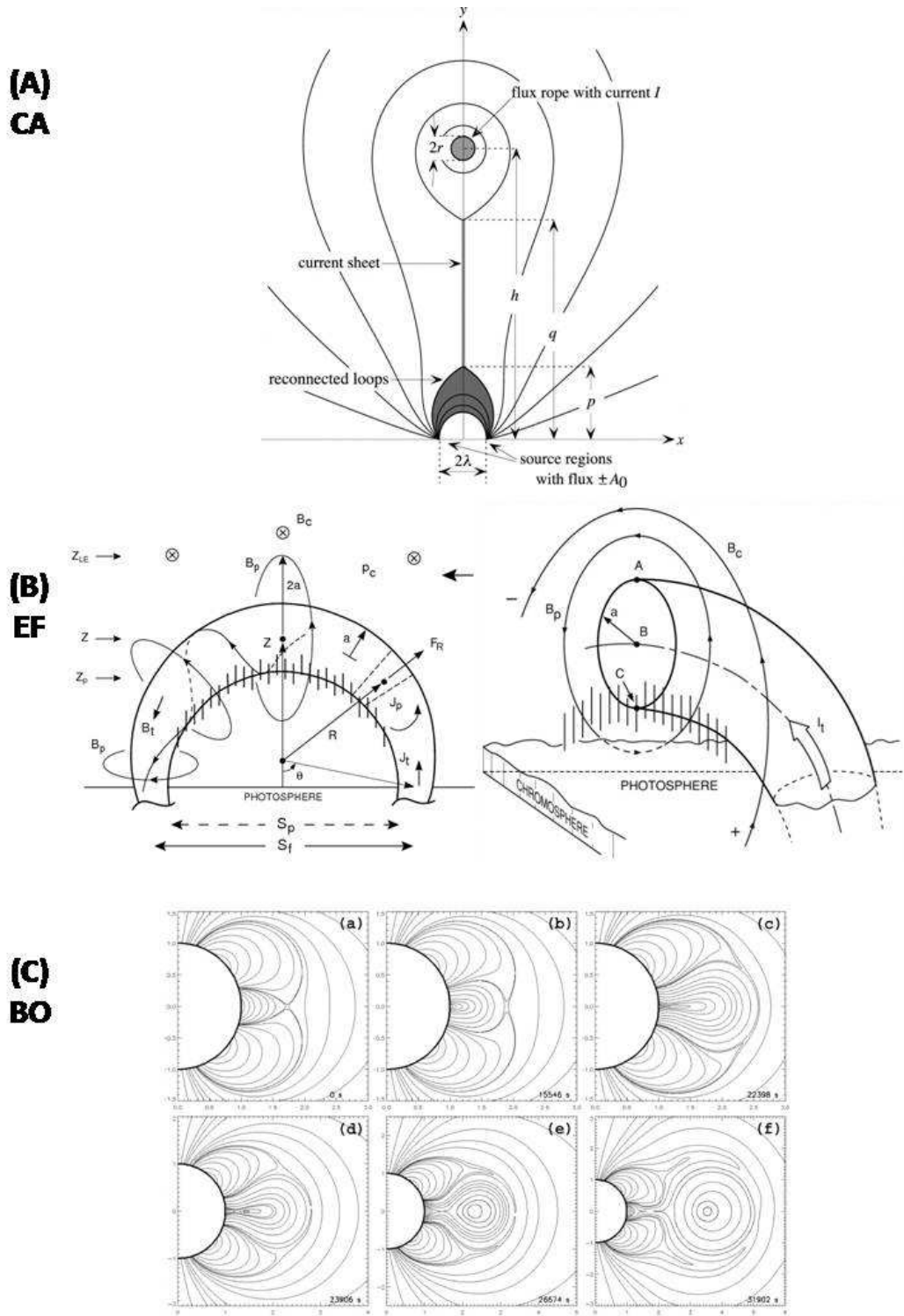}
\end{center}
\caption{{\em (A) CA model}
\citep{LF2000JGR}: the 2D catastrophe model of a CME flux rope.
{\em (B) EF model}:
face-on \citep[left;][]{Chen_etal2006ApJ} 
and cross-section \citep[right;][]{Chen1996JGR}
representations of the EF model.
{\em (C) BO model}:
2.5D simulation results of the BO model to illustrate
the topological evolution of the breakout process \citep{Lynch_etal2004ApJ}.
}
\label{fig:CMEmdls}
\end{figure}

In the model,
the electric current of the flux rope,
composed of toroidal ($I_t$) and poloidal ($I_p$) components,
is assumed to be confined
in a current channel of major radius $R$ and minor radius $a$.
The configuration is illustrated in Fig.~\ref{fig:CMEmdls}(B).
The resulting magnetic field of the flux rope
is also composed of
toroidal ($B_t$) and poloidal ($B_p$) components.
$B_t$ is confined within the current channel
while $B_p$ can extend beyond $r=a$.
The outermost surface of the flux rope is defined at $r=2a$,
which is
where $B_p$ becomes comparable to the ambient coronal magnetic field.
The ambient field is generated by
the current {\em outside} of the flux rope.

The equation governing the radial motion of 
the centroid of
the flux-rope apex is as follows \citep[also see ][]{CK2010ApJ}:
\begin{eqnarray}
M\frac{d^2 Z_{\rm ce}}{dt^2} = \frac{\Phi_p^2}{c^4 L^2 R}
\left[
\ln \left(\frac{8 R}{a}\right) + \frac{1}{2}\beta_p - 
\frac{1}{2}\frac{\bar{B}_t^2}{B_{\rm pa}^2} +
2 \left(\frac{R}{a}\right)\frac{B_c(Z_{\rm ce})}{B_{\rm pa}} -1 + 
\frac{\xi_i}{2}
\right] + F_g + F_d,
\label{eqn:EF}
\end{eqnarray}
where $M$ is the mass per unit length;
$Z_{\rm ce}$ is the height of the flux-rope centroid (ce);
$\Phi_p$ is the poloidal magnetic flux; 
$L$ is the self-inductance;
$R$ and $a$ are the major and minor radius of the flux rope;
$\bar{B}_t$ is the toroidal magnetic field averaged over the minor radius;
$B_{\rm pa}$ is the poloidal field at $r = a$;
$\beta_p \equiv 8\pi(\bar{p} - p_c)/B_{\rm pa}^2$, where
$\bar{p}$ is
the average internal pressure and $p_c$ the ambient coronal pressure;
$B_c$ is the ambient coronal field perpendicular to the flux rope;
$\xi_i$ is the internal inductance;
and $F_g$ and $F_d$ are the gravitational and drag forces.
The motion for
the leading edge (le), 
top of current channel (cr), and the prominence (pr)
are derived
by setting $Z_{\rm le} = Z_{\rm ce} + 2a$, 
$Z_{\rm cr} = Z_{\rm ce}+a$ and $Z_{\rm pr} = Z_{\rm ce}-a$.

Since the minor radius $a$ also changes in time
as the flux rope expands outward,
the following equation is incorporated in the model calculation
to account for the dynamics of the minor radius:
\begin{eqnarray}
M\frac{d^2 a}{d t^2} &=&
\frac{I_t^2}{c^2 a}\left(\frac{\bar{B}_t^2}{B_{\rm pa}^2} -1 + \beta_p\right).
\label{eqn:EF_a}
\end{eqnarray}

To explain CME-associated solar flares and SXR emissions,
\citet{CK2010ApJ} proposed that
the non-zero $d \Phi_p(t)/dt$ would induce an electromotive force (EMF),
$\varepsilon(t) \equiv -(1/c)d \Phi_p(t)/dt$,
which would accelerate charged particles leading to the SXR emissions.
They suggested that an observational indication of such mechanism
would be a coincidence between 
the temporal profile of SXR and $d \Phi_p(t)/dt$.
To test this proposal,
we incorporated 
the comparison between the SXR profile and $d \Phi_p(t)/dt$
in our analysis.

Lastly, 
we compare 
$\partial A(0,y_0)/\partial t$ of the CA model in Eq.~\ref{eqn:recrate1} and 
the $d \Phi_p/dt$ of the EF model.
From the sketch in Fig.~\ref{fig:CMEmdls}(A),
we can see that
in order to satisfy the conservation of magnetic flux,
an increase of the magnetic flux passing through the current sheet 
should be equal to
the increase of the magnetic flux surrounding the flux rope,
which is the poloidal magnetic flux.
In other words,
$\partial A/\partial t$ at the reconnection point
corresponds to $d \Phi_p/dt$ of the flux rope.

\subsection{Breakout model (BO)} \label{sec:mdl_BO}
For a CME to be driven by the BO mechanism \citep{antiochos1999ApJ},
the source region must have a multi-polar magnetic-field configuration,
and must contain a system of magnetic arcades, specifically,
a central arcade, two side arcades and one overlying arcade.
The central arcade is what would later become the CME,
and the shear motion at the foot points is what drives the central arcade
to rise against the overlying arcade.
The process from the rising of a central arcade to 
becoming a CME 
consists of three sets of reconnections:
The first set of reconnection,
occurring between the rising central arcade and
the overlying arcade, removes the confinement,
and is called ``breakout''.
The second set of reconnection cuts off the current sheet 
formed below the rising central arcade,
and allows the CME to escape from the solar surface.
After the central current sheet is cut off,
the side arcades move in,
and form a new current sheet.
The last set of reconnection occurs
in this newly formed current sheet,
and reforms the magnetic fields in the low corona.
The 2.5D topological evolution of this breakout process 
\citep{Lynch_etal2004ApJ} is
illustrated in Fig.~\ref{fig:CMEmdls}(C).
The BO model is calculated by MHD simulations.
The only driver of the instability in the simulation is the shear motion of 
the central arcade \citep{Lynch_etal2009ApJ}.
Neither reconnection rate nor poloidal magnetic flux injection rate
is included in the model.
\citet{Lynch_etal2009ApJ},
however,
found that
the poloidal flux of 
the newly formed flux rope
increases rapidly during the first 1000 sec
of the eruption while the toroidal flux of the same structure remains constant.
This behavior is the same as that in the EF model.
\citet{Lynch_etal2009ApJ}
noted that there was no twist in their system prior to eruption.
Hence, the increase of the poloidal flux was due to the reconnection
converting the overlying field into the poloidal field of the flux rope.
In the BO simulations,
the rate of reconnection depends on the grid-based numerical dissipation.
We can see that 
all three CME models agree that
the eruption is associated to the increase of the poloidal flux of
the erupting flux rope.

The CME kinematics of the BO model have been produced by 
several MHD simulations
\citep{Lynch_etal2004ApJ,Lynch_etal2008ApJ,DVA2008ApJ}.
By tuning the initial conditions,
these simulations generated very different acceleration profiles.
The authors suggested that the BO mechanism can reproduce
different observed CME acceleration profiles by
varying the initial parameters.
Since the BO model simulations are unavailable for our study,
we cannot verify this suggestion,
nor can we make 
accurate and quantitative comparisons between this model
and the trajectories of our two selected CMEs.
Nevertheless,
based on the description of the model,
we can deduce a number of possible observable signatures of 
a CME driven by the BO mechanism:
the source region is multi-polar,
the CME is formed/emerged from an arcade system,
brightenings of source regions and formation of side arcades
occur {\em after} the CME has erupted,
and indications of reconnections can be seen
both before and after the CME lift-off,
and located both above and below the CME flux rope.

\section{CME study} \label{sec:cme}
\subsection{Analysis procedure} \label{sec:cme_analysis}
To examine the dynamics of the CMEs,
we followed the same analysis procedure as described in \citet{lgr2010AA}.
In brief,
we qualitatively and quantitatively compared
the observations with the predictions from the three selected models.
The purpose was, on the one hand, to verify the models and their assumptions, 
and, on the other hand,
to extract the physics of the CMEs from the verified models.
For the qualitative comparison,
we examined various observable properties and phenomena related to the CMEs,
such as the geometry and morphology of the candidate source structures,
SXR and HXR light curves during the process of the events,
and flares and filament eruptions associated with CMEs.
These observed properties and sequence of phenomena were compared with 
the scenarios proposed by the three models to see 
if any of the models is consistent with the observation.
For the quantitative comparison,
different theoretical trajectory profiles were fit to 
the observationally determined trajectory. 
The fitting profiles implemented are
Eq.~\ref{eqn:CA} (CA fit),
$Z_{\rm ce} + 2a$ (EFle fit) and $Z_{\rm ce} + a$ (EFcr fit),
where $Z_{\rm ce}$ and $a$ are determined 
in Eq.~\ref{eqn:EF} and \ref{eqn:EF_a}.

\subsection{Results -- CME03} \label{sec:cme_result3}
CME03 was observed by both SECCHI A and B instruments.
The distance of the CME front from the Sun was determined as follows:
The CME front in each synchronized A- and B- image pair
was traced simultaneously 
using the SolarSoft procedure {\tt scc\_measure.pro}.
The coordinates of the tracing points 
were then used by the procedure to calculate
the true distances of the points from the solar center.
This procedure was repeated ten times at every time step. 
The height of the CME leading edge (LE)
was determined as the average of the ten trial apex heights.
Noticing that the spread of our ten trial heights was generally
comparable to $\sim$1\% of the determined height,
we chose the larger between
1\% of the determined height and
the spread of the ten trial heights
as the observational error.
The velocity and acceleration
were calculated by taking the time derivatives
of the determined height, and 
their respective errors were computed
using the IDL function {\tt derivsig()}.

\begin{figure}
\includegraphics[width=15cm]{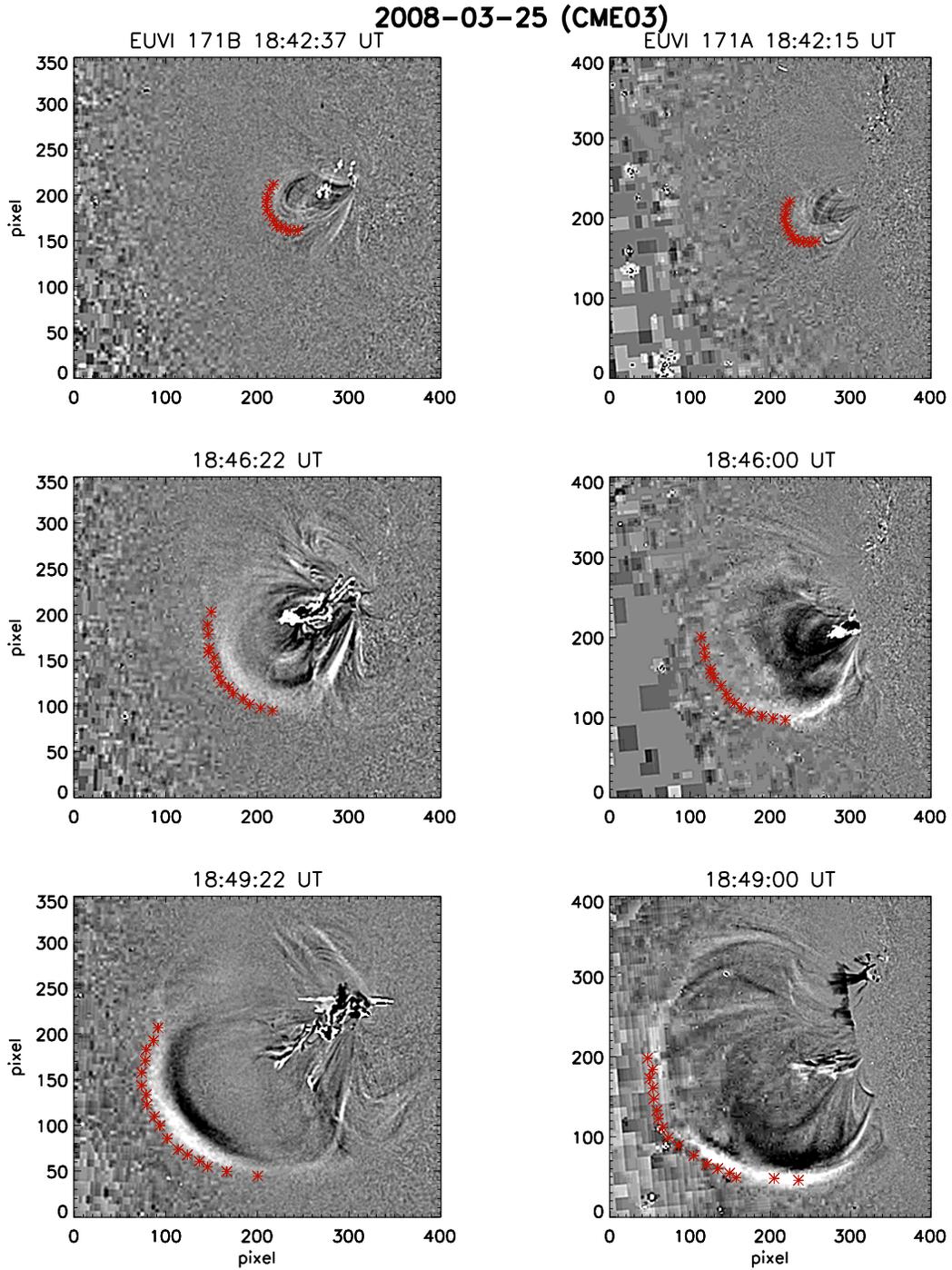}
\caption{Running-difference EUVI images of CME03.
A and B images are plotted in the RHS and LHS panels,
respectively.
The red stars mark the determined CME front.
The observation times are indicated above the figures.
The white rising feature at the core is an erupting prominence.
}
\label{fig:imCME03}
\end{figure}

\begin{figure}
\includegraphics[width=15cm]{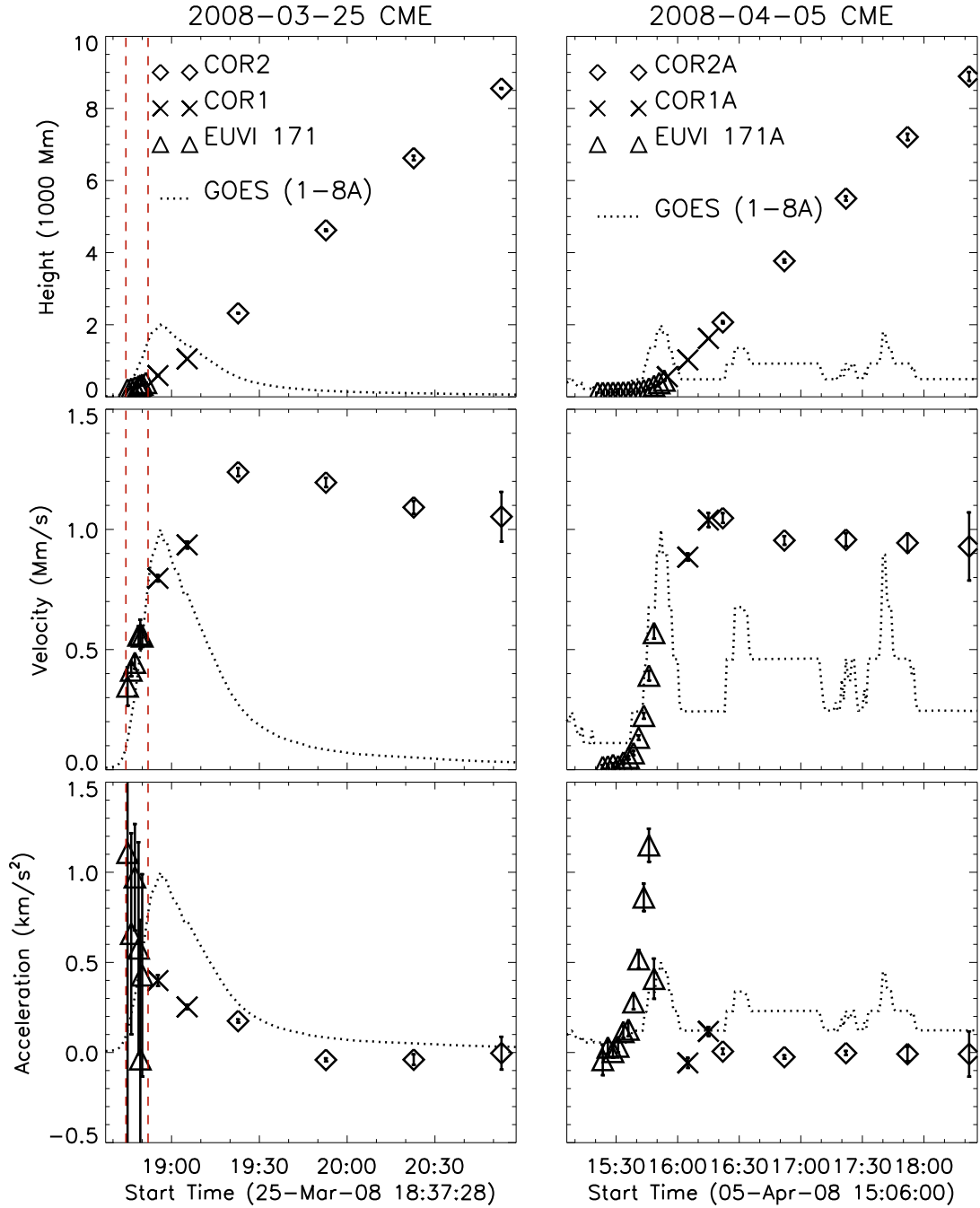}
\caption{The observation-derived kinematics of the CMEs and associated
GOES SXR light curves. 
The results of CME03 and CME04 are presented in the LHS and RHS columns,
respectively.
The magnitude of SXR has been scaled to enhance the visibility.
The red dashed lines mark the beginning and the peak times 
of the HXR emissions associated with CME03.
}
\label{fig:Kin+SXR+HXR}
\end{figure}

Fig.~\ref{fig:imCME03} shows the initial eruption of this CME.
The running-difference images of
the EUVI-B and A instruments are shown in
the left-hand-side(LHS) and right-hand-side(RHS) columns respectively, and
the observation times are as indicated.
The red stars mark the CME LE.
The EUVI images have been enhanced by a wavelet image-processing technique
\citep{SVH2008ApJ}.
The bright feature at the CME core is an erupting prominence.
From the EUVI images of the source region during the entire CME process,
we did not find indications of a BO process,
such as multiple reconnections and reformation of
multiple arcade systems.
An enhanced HXR emission in the
RHESSI 18--30~keV energy band was detected from AR10989 during this CME event.
\citet{Temmer_etal2010ApJ} deduced from the RHESSI HXR image
that the main emission was from one of the foot points of the CME.
The profiles of the SXR light curve and the CME kinematics are plotted
in the LHS column of Fig.~\ref{fig:Kin+SXR+HXR}.
The height, velocity and acceleration are shown in
the top, middle and bottom panels, respectively.
The SXR emission magnitude has been scaled for clarity. 
The two vertical dash lines mark the beginning and peak times
of the HXR emission.
We found that
the determined initial LE is higher than 150~Mm
and the initial velocity is greater than 300~km/s,
suggesting that the very beginning of the eruption 
was not detected in our observation.
\citet{Temmer_etal2010ApJ}
detected the flux rope at an earlier time and lower height.
Their kinematics plot showed that the velocity reached approximately 300~km/s
and acceleration was $\sim 0.8$ km/s$^2$ at $\sim$~18:44~UT,
which is consistent with our {\bf values (cf. Fig.~\ref{fig:Kin+SXR+HXR}).}

\begin{figure}
\includegraphics[width=15cm]{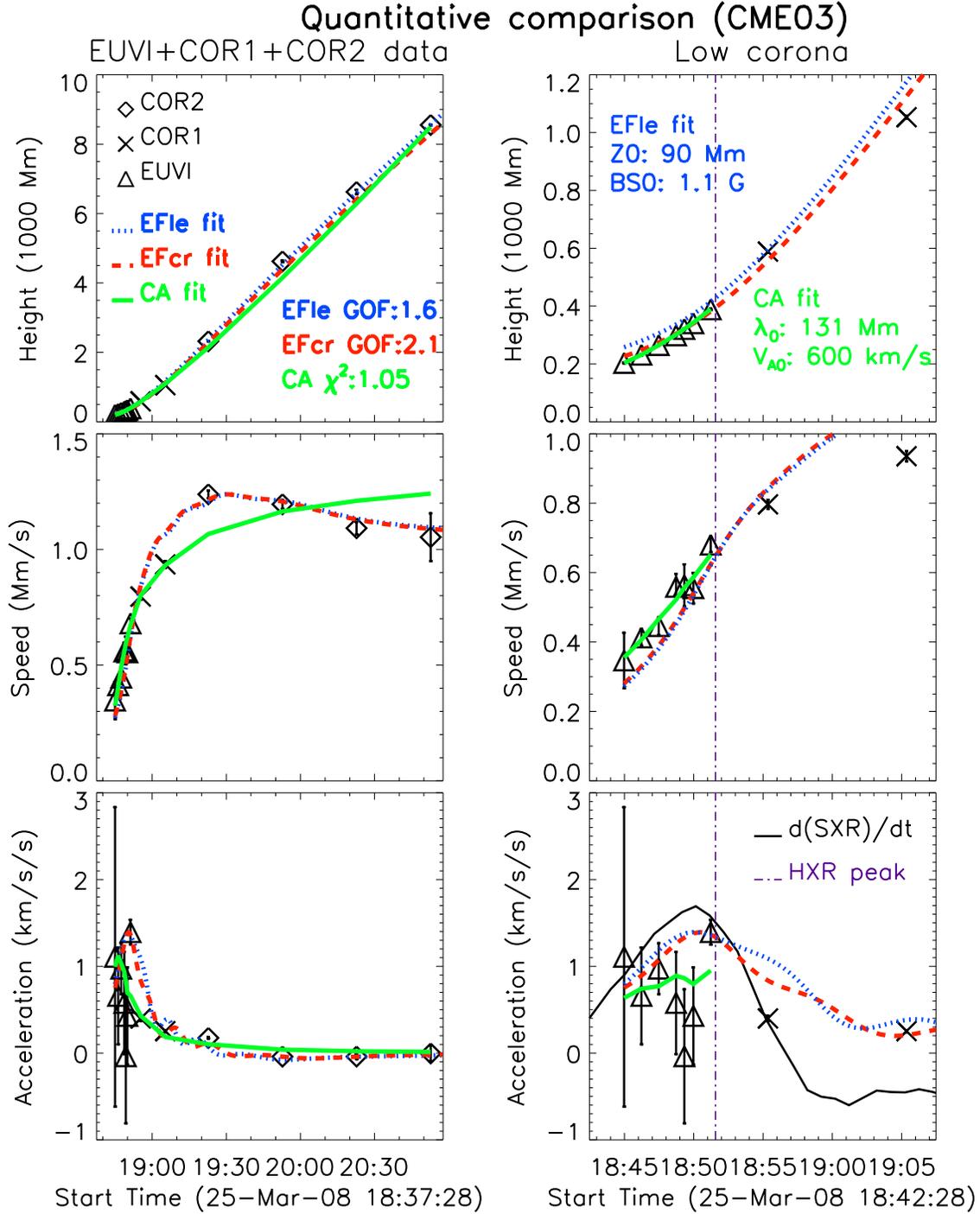}
\caption{The fitting results for CME03.
The data and best-fit results are plotted in symbols and lines as indicated.
The LHS column shows the entire process 
while the RHS shows only the low-corona part
to enhance visibility of the initial eruption.
The dot-dash line in the RHS column marks the time of HXR emission peak.
The solid curve in the lower right panel is the
time derivative of SXR light curve.
{\it EF fit:}
Z0 is the initial equilibrium height from the bottom of corona;
BS0 is the external coronal field perpendicular to the flux rope at Z0.
{\it CA fit:}
$\lambda_0$ is
the height where loss-of-balance happens; 
$V_{\rm A0}$ is
the Alfv\'en speed at $\lambda_0$.
}
\label{fig:fit_cme03}
\end{figure}

Our fitting results are presented in
Fig.~\ref{fig:fit_cme03}.
The data are plotted in symbols with the error bars
and the best-fit results
in continuous lines,
as denoted in the figures.
To allow detail inspection of
both the entire kinematics profile and
the initial-phase profile, 
the results of the entire set of data are placed in the LHS column
and those of the low corona part in the RHS column.
The purple dot-dashed vertical line in the RHS column marks the time of the
RHESSI HXR peak emission.
The black solid curve 
in the bottom right panel 
is
the time derivative of the GOES SXR light curve.

The results in the RHS show that both EF and CA models match
the observationally determined kinematics 
in the low corona 
reasonably well.
However,
the CA fit deviates from the data points after $\sim$19:10~UT
while the EF continues to follow the observation up to 
the last observed point.
This indicates two things:
one is that
the EF formulation is valid through the entire CME process,
and
the other is that
under the formulation of the CA model, 
the ideal MHD process becomes inappropriate after the initial eruption,
and magnetic reconnections must occur to accelerate the CME
to reach the observed speed.
To obtain an estimate of the reconnection rate,
we use 
the same computation algorithm and model configuration
as implemented in \citet{LF2000JGR}
except that the source-region magnetic field strength 
is set to 40 Gauss, 
which is approximately the value averaged over a $15^{\circ}$ patch
around AR10989,
and then tune the Alfv\'en Mach number $M_A$ until
the code produces
a kinematic profile that qualitatively resembles the observational one.
Since the plasma condition and the configuration of the flux rope 
are built-in to the code and un-adjustable,
the comparison between the computed profile and the observation
is only qualitative.
$M_A(\equiv V_R/V_A)$, where $V_R$ is the plasma inflow speed
(cf. Sec.~\ref{sec:mdl_CA}),
thus obtained is $\sim 0.06$.
Note that this value is obtained for a model flux-rope
as specified in \citet{LF2000JGR},
and may not represent the actual reconnection rate.

\begin{figure}
\includegraphics[width=15cm]{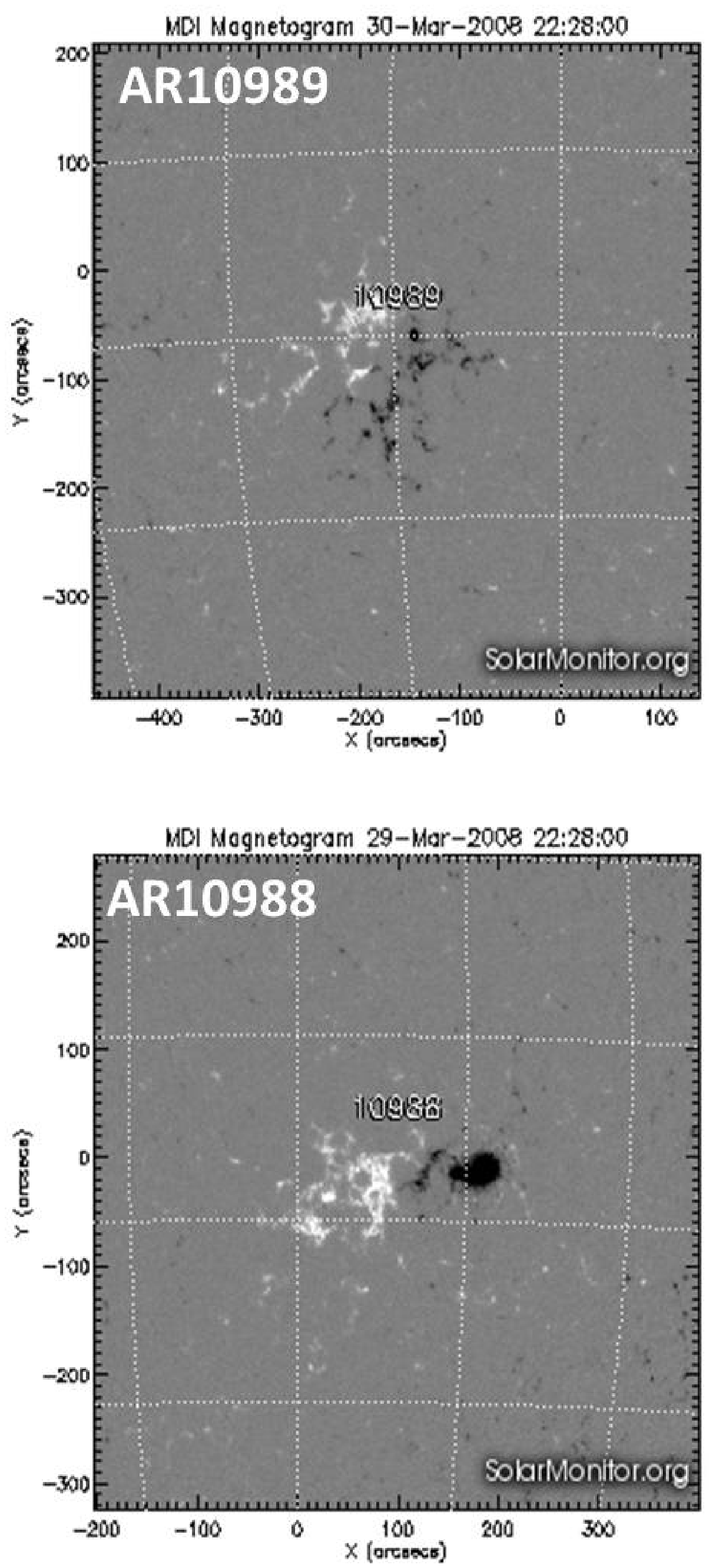}
\caption{MDI magnetograms of AR10989 (upper panel)
and AR10988 (lower panel) when they are at the disk center.}
\label{fig:mdi_10988-9}
\end{figure}

The critical height $\lambda_0$ 
(i.e., the height when the loss of balance happens)
and the Alfv\'en speed at this height
obtained from the best CA-fit
are $\lambda_0 \approx 131$~Mm and $V_{\rm A}(\lambda_0) \approx 600$~km/s.
These values correspond to a foot-point separation of $\approx 262$~Mm,
and an ambient magnetic-field strength of $\approx 22$~G
for a plasma density $\approx 10^{-4}$ kg km$^{-3}$.
From the best EF-fit result,
the initial height of the flux rope ($Z0$) is approximately 90~Mm 
measured from the base of corona,
and the foot-point separation, $S_f$,
is $\approx 180$~Mm.
The MDI magnetogram of AR10989 at the disk center 
(cf. upper panel in Fig.~\ref{fig:mdi_10988-9}),
shows 
that the size of the region is approximately 200 arcsec 
($\approx 150$~Mm).
Since the flux-rope foot points cannot be directly observed,
we use this distance as an observational proxy for 
the actual foot-point separation.
Hence,
the foot-point separation predicted from the best CA- and EF-fit
are respectively 75\% and 20\% greater than 
a rough estimate based on the size of the region from MDI data.
In this comparison, the EF model agrees with the observation
better than the CA model does.

Comparison between EFle (blue dotted line) and EFcr (red dashed line)
in the top right panel
shows that
EFcr fits 
the EUVI part of the data
slightly better than EFle does,
implying that 
the LE identified and tracked in the EUVI 171\AA~images
and in the COR1/COR2 images, which observe in white light, 
might not correspond to the exact same feature.
The bottom right panel demonstrates a temporal coincidence between
the peak of $d{\rm I}_{\rm SXR}/dt$(solid line) and the acceleration peak of
the EFle fit (blue dotted line).
The peak HXR emission occurs slightly later.
This slight delay can also be seen in the results by 
\citet{Temmer_etal2010ApJ} 
(cf. Fig.~$12$ in the paper).

\begin{figure}
\includegraphics[width=15cm]{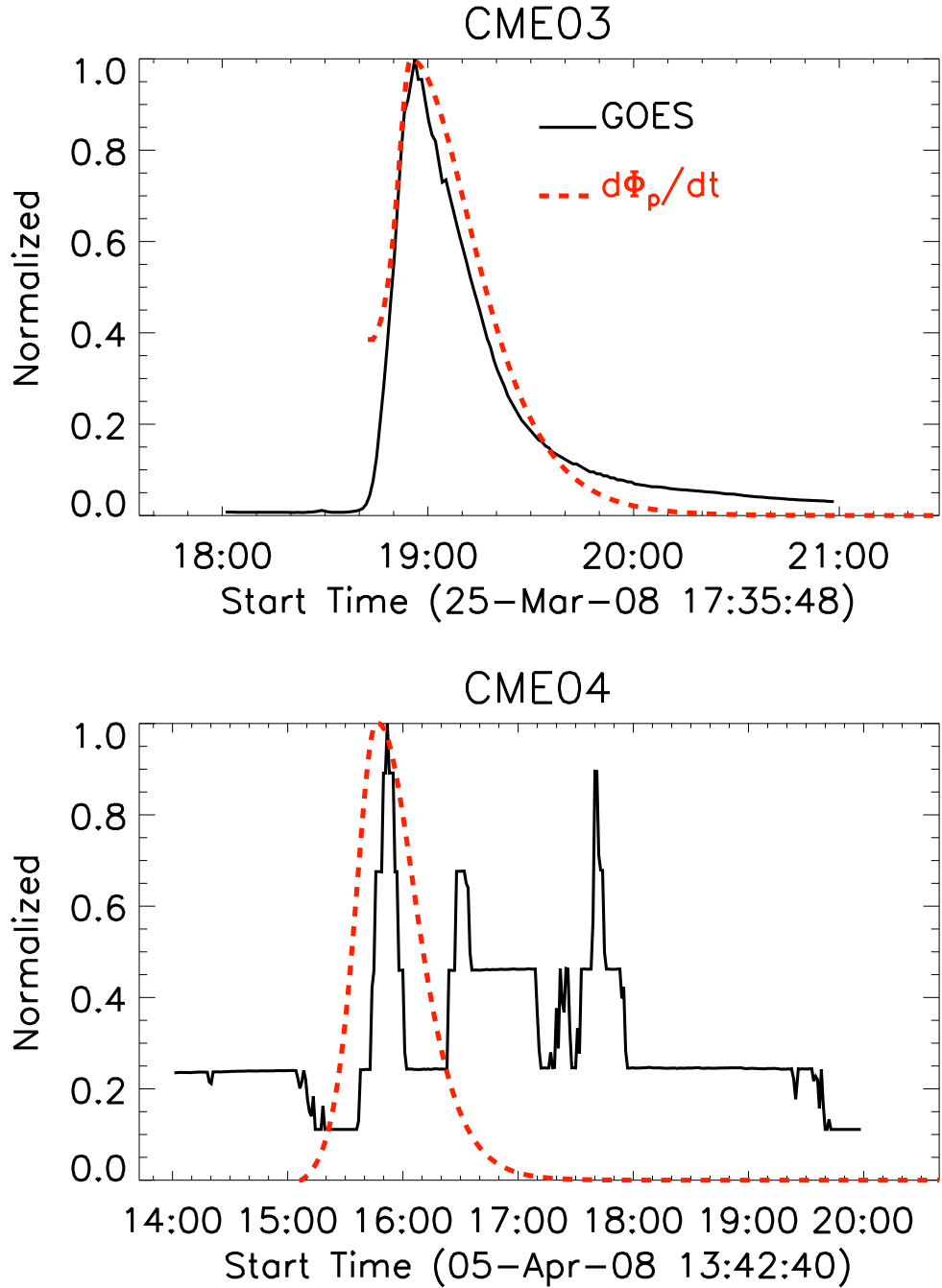}
\caption{Comparison of the GOES SXR profile (solid black line) and 
the poloidal magnetic flux injection rate, $d\Phi_{\rm p} / dt$, 
(red dashed line) predicted by the EF best-fit.
The results for CME03 and CME04 are shown in the upper and lower panels, 
respectively.
All magnitudes have been normalized to allow easy comparison of the shapes.
}
\label{fig:SXRdPhi}
\end{figure}

The SXR emission light curve (I$_{\rm SXR}$)
and the predicted temporal form of $d\Phi_{\rm p}/dt$ from our best fit
of this event 
are compared in the upper panel of Fig.~\ref{fig:SXRdPhi}.
The two curves coincide with each other,
and the timing of the peaks agree with each other very well.
The lack of the initial part of $d \Phi_{\rm p}/dt$
reflects the absence of the data for the initial rising stage
in our observational data.
We note that 
the decaying phases of I$_{\rm SXR}$ and $d \Phi_{\rm p}/dt$ differ.
Because the EF model does not include radiation or radiation-plasma interaction,
it is beyond the scope of this paper to address the decay time scale.
In addition,
because the model does not include particle energization/acceleration
underlying the SXR emissions,
it produces no prediction of the energy spectrum.
It does, however,
directly predicts the temporal profile of the electromotive force
($\varepsilon(t) \equiv -(1/c)d \Phi_p(t)/dt$)
that
can accelerate particles.
Thus,
the comparisons here are between the predicted and observed temporal profiles
excluding the details of decay phase.
The detailed manifestations and magnitude of $d\Phi_p/dt$ cannot be resolved
or measured at this time,
and must await future observations.

\subsection{Results -- CME04} \label{sec:cme_result4}
CME04 was seen as a west limb event from the viewpoint of 
the STEREO-A instruments, but
was occulted by the west limb of the Sun in the EUV images seen 
from the STEREO-B.
Therefore, 
the kinematics for this event was derived solely from the STEREO~A instruments.
The motion of the CME was determined by tracing
the CME front in each image.
The procedure was repeated ten times for each image 
to obtain an average height.
Assuming that the CME propagates radially from the eruption site,
we corrected the projection effect by
the trigonometric formula
\citep[e.g.,][]{CK2010ApJ}:
\begin{equation}
h = R_{\rm obs}\frac{\sin \alpha}{\sin(\alpha + \mu)},
\end{equation}
in which $h$ is the corrected height from the solar center,
$R_{\rm obs} = 214 R_\odot$ is the observer--Sun distance,
$\alpha$ is the elongation,
and $\mu \equiv \cos^{-1} (\cos \phi \cos \theta)$, where
$\theta$ and $\phi$ are the heliocentric longitude and latitude
of the source region from the STEREO-A viewpoint.
As described in Sec.~\ref{sec:cme_result3},
the observational height error was either
the standard deviation of the ten trials
or 1\% of the determined height,
whichever is larger.
The observational velocity, acceleration,
and their respective errors were subsequently derived
from the determined height and errors 
as described in Sec.~\ref{sec:cme_result3}. 

\begin{figure}
\includegraphics[width=15cm]{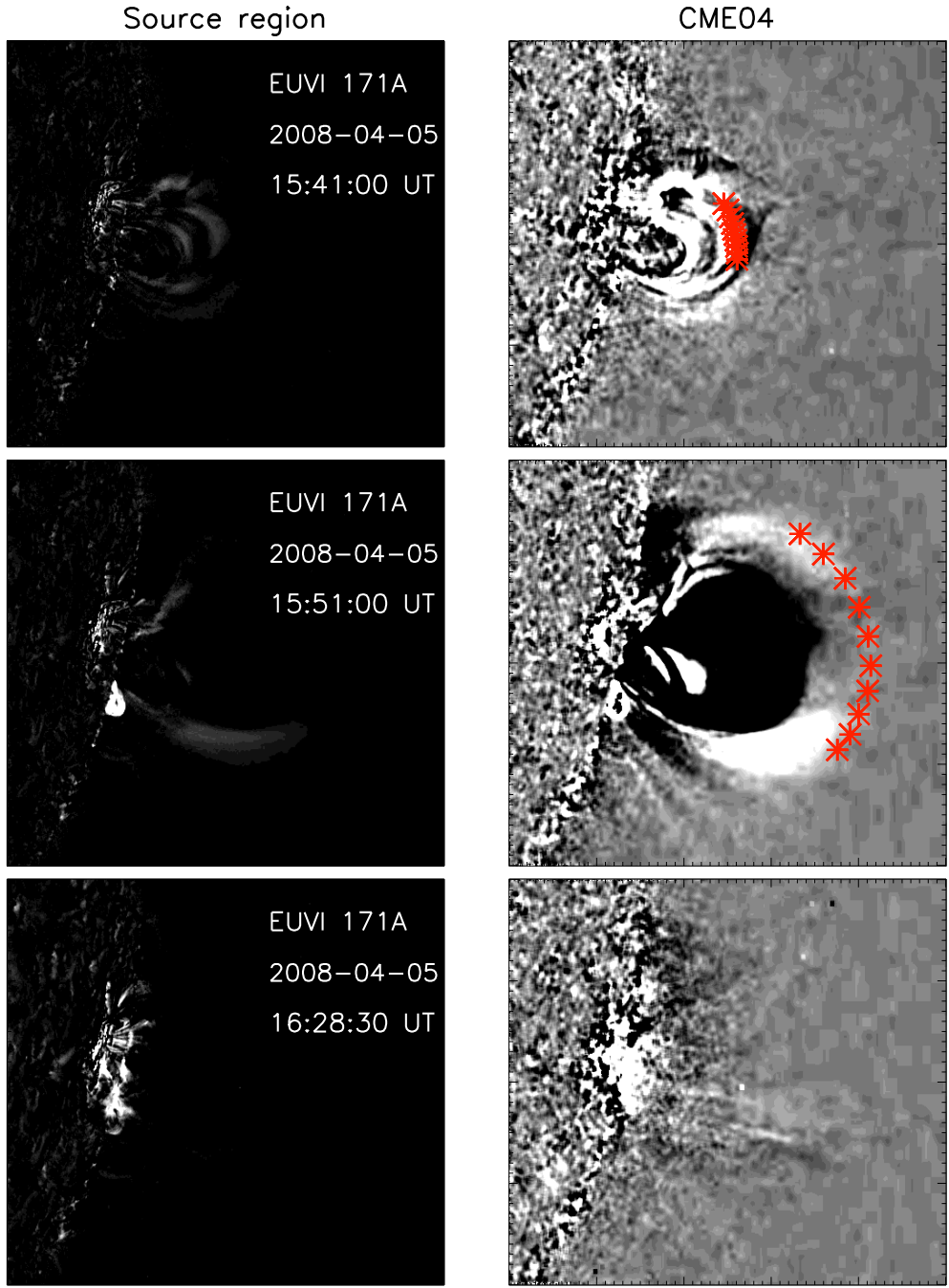}
\caption{Simultaneous evolution of CME04 
and its source region (AR10988) at selected times.
The LHS column shows the source region,
and the RHS are the corresponding running difference images to show CME04.
The observation times are indicated in {\bf the images on the LHS}.
The source region was originally quiet in EUVI 171 \AA, 
but a transient side arcade (middle panel) and 
multiple brightenings (bottom panel)
later appeared during the CME process.
}
\label{fig:imCME04}
\end{figure}

The source region evolution during the process of CME04 is
shown in the LHS column of Fig.~\ref{fig:imCME04},
and the running difference images of the respective images
are plotted in the RHS column to 
show the CME process at the same time.
The red stars mark the CME LE.
The observation time of the images in the same row
is indicated in the LHS panels.
The middle row reveals 
the appearance a bright small-scale side arcade at one foot point 
of the CME
when the CME reached the boundary of the EUVI field of view.
The last row shows brightening of
several foot points at the source region
after the CME flew off.
The formation of the bright side arcade and 
brightenings at the source region 
after the eruption
are part of the features expected by the BO model
(cf. Sec.~\ref{sec:mdl_BO}).
However,
without information from a quantitative comparison,
we cannot confidently verify the physical consistency.

The kinematics of this CME and the associated SXR light curve 
are shown in the right column of Fig.~\ref{fig:Kin+SXR+HXR}.
The height, velocity and acceleration of the CME 
are plotted as symbols in the top, middle and bottom panels, respectively.
The magnitude of the SXR emission has been scaled for clarity.
In contrast to CME03,
CME04 was indeed observed from its initial equilibrium state,
as evidenced by the near-zero initial speed.
The observed SXR light curve shows several small emission enhancements.
We caution that
the light curve might not represent the actual SXR emission profile because
the source region was occulted by the west limb of the Sun
from the view angle of GOES.
Comparing the timing of the SXR peaks with the times indicated in 
Fig.~\ref{fig:imCME04},
we found that the first peak ($\sim$~15:51~UT) coincided with 
the appearance of the side arcade
and the second one ($\sim$~16:30~UT) with 
the brightenings at the source region,
indicating a correlation between the two SXR peaks and 
the brightening of the arcade and source region.

\begin{figure}
\includegraphics[width=15cm]{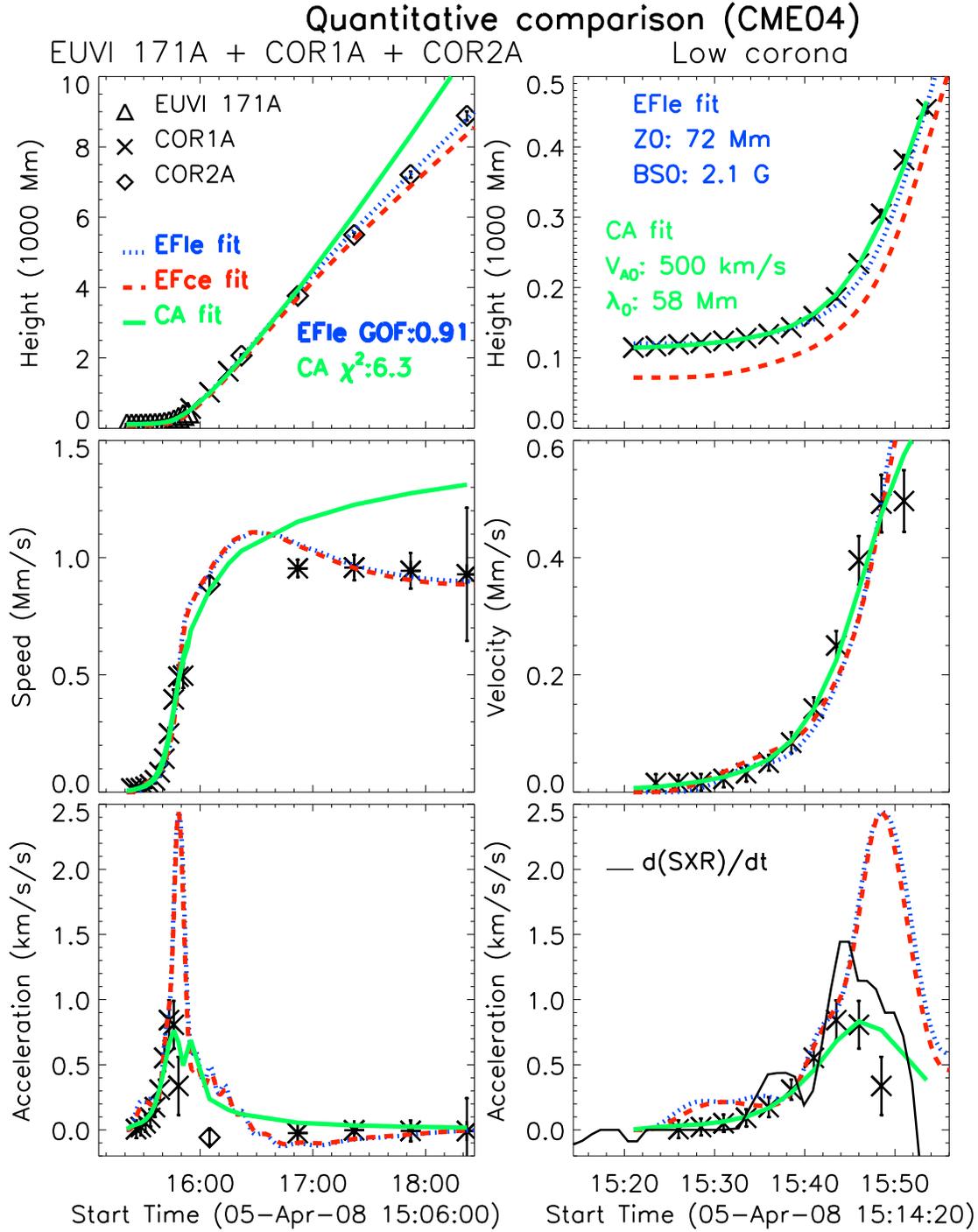}
\caption{EF and CA fitting results of CME04.
The height, velocity and acceleration are plotted in
the top, middle and bottom panels, respectively.
The symbols and line styles are as described in Fig.~\ref{fig:fit_cme03}.
}
\label{fig:fit_cme04}
\end{figure}

The quantitative examination of CME04 is presented in
Fig.~\ref{fig:fit_cme04}.
The symbols and the line styles are the same as those in
Fig.~\ref{fig:fit_cme03}. 
The CA fit (green solid line) is the best match to
the EUVI part of the kinematics,
but deviates from the observation results in COR2.
The deviation, as explained in Sec.~\ref{sec:cme_result3},
indicates that magnetic reconnection is required to produce
the observed CME process under the CA formulation.
To estimate $M_A$ at the reconnection point,
we follow the same procedure as described in Sec.~\ref{sec:cme_result3},
and obtained $M_A \approx 0.025 - 0.03$ for 
source-region magnetic field $B = 80$~G,
which is approximately the value averaged over a $15^{\circ}$ patch
around AR10988.

The best CA fit predicted the critical height $\lambda_0$ 
of this event to be 58~Mm
and the Alfv\'en speed at this height to be 500~km/s.
These values translate to a foot-point separation of $\approx 116$~Mm
and an ambient magnetic field strength of $\approx 18$~G for
a plasma density of $\approx 10^{-4}$~kg~km$^{-3}$.
For the EFle fit (blue dotted line),
it over predicted the peak acceleration in the low corona (15:46 -- 15:54~UT),
but closely follows the rest of observationally determined kinematics.
The initial height and the foot-point separation
obtained from our EF best fit are
$Z0=72$~Mm and $S_f=140$~Mm, respectively.
The foot-point separation {\bf values of EF and CA fits 
are both} consistent with the size
of AR10988 at photosphere ($\sim 150$~Mm),
as estimated from the MDI magnetogram 
(cf. lower panel in Fig.~\ref{fig:mdi_10988-9}).
The required $d\Phi_{\rm p}/dt$ and the observed SXR of this event
are compared in the lower panel of Fig.~\ref{fig:SXRdPhi}. 
Since the source region of the event was behind the solar limb from 
the view point of GOES,
the instrument can only detect the later part of the SXR emission
when the SXR loops had grown sufficiently high to be seen
over the limb.
Therefore,
the SXR signal should be detected later than $d\Phi_p/dt$.
The plot shows that, as expected,
the SXR peaked later than $d\Phi_{\rm p}/dt$.
The width of the SXR profile is, however, much narrower than
the width of $d\Phi_p/dt$.
Since $d\Phi_p/dt$ is the main driving mechanism in the EF model,
the wider $d\Phi_p/dt$ profile and the larger peak acceleration 
predicted by the model
imply that 
some assumptions in the model,
such as the proposed SXR emission and/or CME driving mechanisms,
are not suitable for this event.
We should point out that
the occultation of the SXR emission by the solar limb and
the fact that the EF model does not include 
radiation mechanism
inevitably contribute to the discrepancy.
\section{Helioseismic study of subsurface structures} \label{sec:sub}
To investigate possible association between eruptivity
and subsurface structure of an active region,
we applied local helioseismic inversion techniques 
\citep{Hill1988ApJ,Basuetal2004ApJ}
to probe the subsurface structures of six ARs,
and consulted the following resources 
for the information of relevant flare and CME activities:
{\tt http://www.solarmonitor.org}, NOAA/USAF Active Region Summary, and
{\tt http://cor1.gsfc.nasa.gov/catalog/}.
In the following, 
we give a brief description of the principle of helioseismic inversion.

The turbulence in the near-surface layer of the Sun excites pressure waves
that propagate through the solar interior at the sound speed.
Since the sound speed increases with temperature,
the waves are totally refracted at specific depths according 
to their wave lengths.
The longer the wave length, the deeper they can penetrate.
The refracted waves travel upward, and are reflected at the surface
due to the large density drop across the surface.
In other words,
these waves travel through and resonate between the surface
and a specific depth.
They manifest themselves as Doppler shifts at the solar surface,
of which the observed patterns can be
decomposed into different spherical harmonic modes.
The observed frequencies of these pressure modes (p-modes)
depend on the structure and dynamics of the layer inside the Sun
they travel through.
Helioseismic inversion is a procedure
to infer the unobservable internal structure and dynamics
from these observable frequencies.
The application of the inversion procedure to localized regions 
on the solar surface is called local helioseismic inversion.
The purpose of
local helioseismic inversion is to reveal local non-uniformity,
such as flow patterns or structural differences.
In this study, the subject of interest is
the latter, the difference in the structural properties.

\subsection{Helioseismic inversion method} \label{sec:sub_inv}
The starting point of helioseismic  inversions  is the
linearization of the oscillation equations around a known solar model 
(the so-called reference model) using the variational principle.
The frequency differences can then be related to the differences
in the structural properties.
Since p-modes are acoustic waves,
the adiabatic sound speed ($c_g \equiv \sqrt{\Gamma_1 P/\rho}$)
and density ($\rho$)
are natural choice of the thermal structure properties for the inversion.
The  relation between the differences in frequency and
these two variables ({\it i.e.}, $c$ and $\rho$)
can be written as \citep[e.g.,][]{DPS1990MNRAS}:
\begin{eqnarray}
\frac{\delta \omega_i}{\omega_i} &=&
\int_0^R K^i_{c_g^2, \rho}(r) \frac{\delta c_g^2}{c_g^2}(r) {\rm d}r +
\int_0^R K^i_{\rho, c_g^2}(r)
     \frac{\delta \rho}{\rho}(r) {\rm d}r +
\frac{F_{\rm surf}(\omega_i)}{E_i} + \epsilon_i,
\label{eqn:inv}
\end{eqnarray}
where
$K^i$ are the  kernels derived from the reference model,
$\delta\omega_i/\omega_i$ is the relative frequency difference of the $i$th
mode,
$\epsilon_i$ is the observational error in $\delta\omega_i/\omega_i$, and
$F_{\rm surf}(\omega_i)/E_i$, usually called the ``surface term'',
represents the effect
of uncertainties in the model close to the surface.
Here, $E_i$ is a measure of the mode inertia.
Other pairs of variables such as ($\Gamma_1$, $\rho$) can be used instead of
the ($c_g^2$, $\rho$) pair used above.
$\Gamma_1$ is the adiabatic index defined as $(\partial \ln P/\partial \ln \rho)_s$,
where $s$ is entropy.

\subsection{Analysis procedures} \label{sec:sub_mag}
To examine the subsurface structure of active regions,
we follow the strategy as described in \citet{Basuetal2004ApJ},
who used
the frequency differences between
the target active region and a reference quiet Sun (QS) region,
instead of between active regions and a solar model.
The reference QS region is selected to be at a same latitude as 
the target AR,
and both of them are tracked 
over a time interval centered at their central meridian passages.
By doing so,
we can reduce the systematic effects of projection on the frequencies
and also minimize
some of the effects of non-adiabaticity in the
near-surface layers that are not considered in solar models.
The level of surface magnetic activity 
is characterized by
the Magnetic Activity Index (MAI) calculated from the MDI magnetograms,
as described in \citet{Basuetal2004ApJ}. 
The difference 
in MAI, $\Delta$MAI, between each AR and its reference QS,
along with
the tracking time of each target AR and
the coordinates of each AR-QS pair,
are listed in Table~\ref{tab:tab_coord}.

The mode frequencies are determined by the ring diagram analysis 
\citep{Hill1988ApJ}.
The ring diagrams are three-dimensional power spectra of the Doppler shifts
in small patches
of the solar surface tracked at the Carrington rate.
To obtain the frequencies of different modes,
we fitted the power spectrum using a model spectral profile
in the manner described in \citet{Basuetal2004ApJ}.
The fit parameters are then interpolated
to integer $n,\ell$ pairs,
where $n,\ell$ are radial and angular mode numbers of spherical harmonics.
The frequency differences between AR and QS are subsequently inverted
to obtain the structural properties based on the inversion equation
Eq.~\ref{eqn:inv}.
The inversion was performed by the computational technique,
Subtractive Optimally Localized Averages \citep[SOLA;][]{PT1992AA}.

The current standard solar model,
from which the inversion kernels are derived,
does not include magnetic fields
because
the gas pressure inside the Sun is several orders of magnitude higher 
than the magnetic pressure.
The study by \citet{lbl2009SoPh} showed that,
even in the near-surface layer below active regions,
where the gas pressure is low compared to other parts of the interior,
the magnetic pressure averaged over volume is nearly 100 times 
smaller than the gas pressure.
In other words, 
the effect of the magnetic fields on the structure is considered to be
sufficiently small to justify the variational principle.
However,
while the kernel errors only contribute to the higher order terms
according to the variational principle,
the quality of the current helioseismic data has become 
capable of discerning such higher order effects.

The presence of magnetic fields can affect the frequencies of waves
in two ways:
Firstly, the magnetic fields change the thermal structure
of the medium that
the waves travel through, which, in turn, changes the frequencies of the waves.
Secondly, the plasma waves are directly affected by the magnetic fields
through the Lorentz force.
The modification to the frequencies results from
both structural and non-structural ({\it i.e.}, Lorentz force)
effects of the magnetic fields.
Since the two effects are inseparable in the observed frequencies,
the ``structures'' revealed by the inversions could partly 
be due to
the frequency difference caused by Lorentz force on wave propagation.
By using solar models that include magnetic fields
and inversion kernels computed from non-magnetic reference models,
\citet{lbl2009SoPh}
showed that the ``sound'' speed revealed by the inversions
in the presence of magnetic fields
is in fact 
a combination of adiabatic sound speed
($c_g \equiv \sqrt{\Gamma_1 P_{\rm gas}/\rho}$) and
Alfv{\'e}n speed ($c_A \equiv B/\sqrt{4\pi\rho}$).
To distinguish this property from $c_g$,
we use $c_T$ {\bf ($\equiv \sqrt{\Gamma_1 P_T/\rho}$)}
to represent 
the speed obtained from the inversions,
where $P_T(=P_{\rm gas}+P_{\rm mag})$ is the total pressure
and $P_{\rm mag} = B^2/8\pi$ is the magnetic pressure.
Hence, the relation between $c_T$, $c_g$ and $c_A$ is 
$c_T^2 = c_g^2 + \Gamma_1 \; c_A^2$.
Note that the helioseismic inversion speed $c_T$ is to be distinguished from
the magnetosonic speed $c_M^2 = c_g^2 + c_A^2$.
While the
speed of sound is directly related to temperature,  there is
no simple, direct relationship between the inversion speed $c_T$ and
either temperature or magnetic fields.
Searching for an observable representation of 
the subsurface magnetic structure,
\citet{lbl2009SoPh} found 
an asymptotic linear dependence
between
$\delta \Gamma_1/\Gamma_1$
and
$\delta \beta_H \equiv \delta (P_{\rm mag}/P_{\rm gas})$
at every depth. 
The relation is different at different depths.
They empirically derived a linear mathematical expression for
each selected depth within the region where the inversion is reliable,
and used the relation to infer $\delta \beta_H$.

$\beta_H$ represents the relative importance 
between the magnetic-field and thermodynamical effects.
The effect from a strong magnetic field would still be un-detectable
if the gas pressure is much larger than the magnetic pressure.
Large $\beta_H$ 
means the effect from the magnetic field is large,
which can result from 
either low gas density or strong magnetic field or both.
In this study, 
we use $\delta \beta_H$ to infer
the subsurface magnetic structural difference
between an AR and its reference QS region.

To probe the subsurface temperature differences,
we rewrite $\delta c_T^2/c_T^2$ as follows:
\begin{eqnarray}
\delta c_T^2/c_T^2 &=& \delta \Gamma_1/\Gamma_1 + 
\frac{\delta (P_T/\rho)}{P_T/\rho}.
\label{eqn:dc2dg1}
\end{eqnarray}
We can see that the difference between $\delta c_T^2/c_T^2$
and $\delta \Gamma_1/\Gamma_1$
can reflect 
the difference in 
temperature.
We point out that the temperature difference is caused by
a combination of the differences in magnetic and gas pressure.
In short,
the subsurface magnetic structure in our analysis was inferred from
$\delta \Gamma_1/\Gamma_1$ and $\delta \beta_H$,
and the subsurface temperature profile
from $\delta c_T^2/c_T^2-\delta \Gamma_1/\Gamma_1$.

\subsection{Results and discussion} \label{sec:sub_results}
\begin{figure}
\includegraphics[width=15cm]{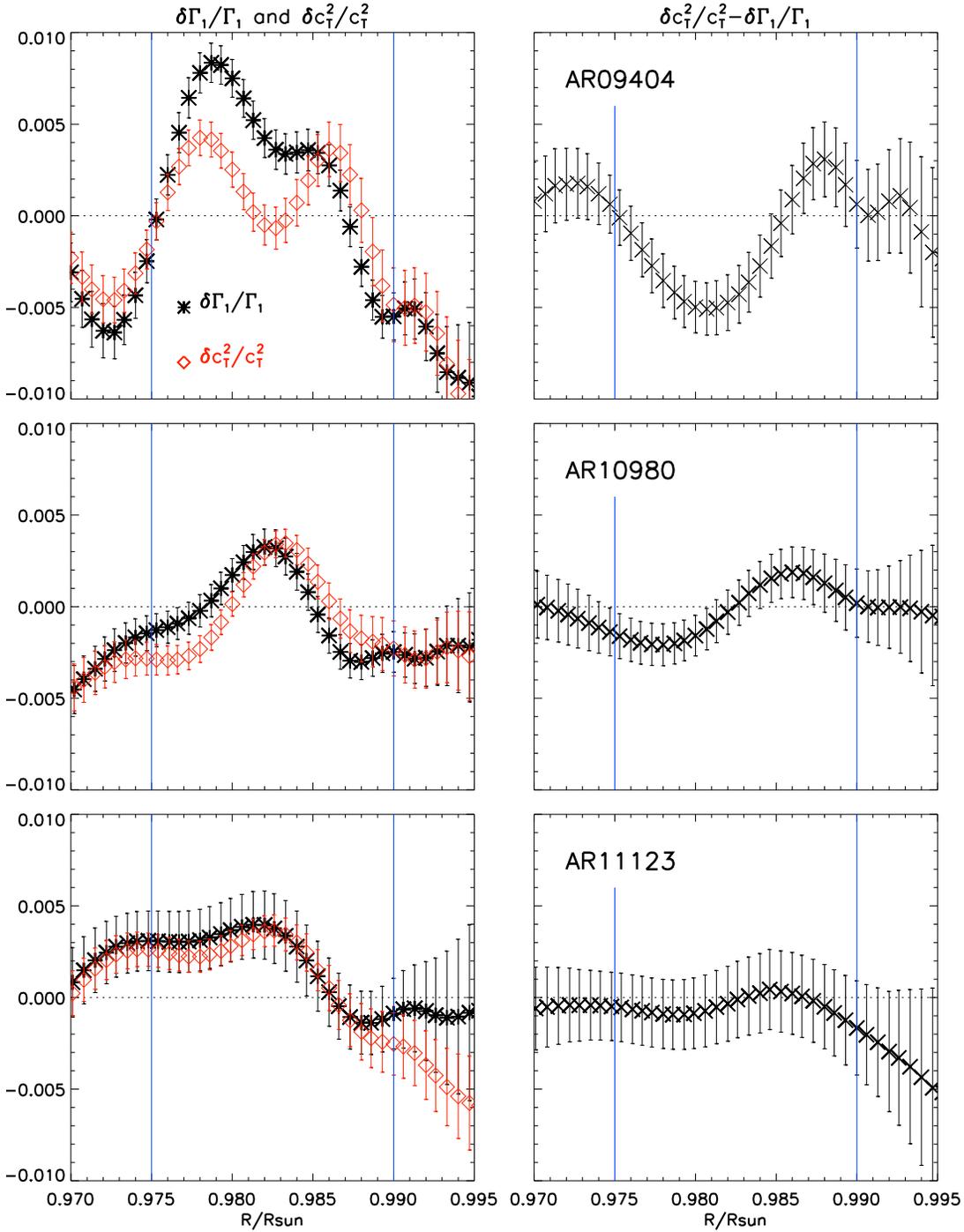}
\caption{
The inversion results of AR09404, 10980 and 11123.
The results in the same row correspond to the same active region,
as indicated in the right panels.
The regions are arranged from the least eruptive on the top row
to the most eruptive on the bottom.
The LHS column shows
$\delta c_T^2/c_T^2$ (red diamond) and $\delta \Gamma_1/\Gamma_1$ (black star).
The RHS column shows 
$\delta c_T^2/c_T^2 - \delta \Gamma_1/\Gamma_1$.
The two blue vertical lines are to mark the region of interest,
which is $0.975-0.99R_\odot$.
}
\label{fig:dg1dc2a}
\end{figure}

\begin{figure}
\includegraphics[width=15cm]{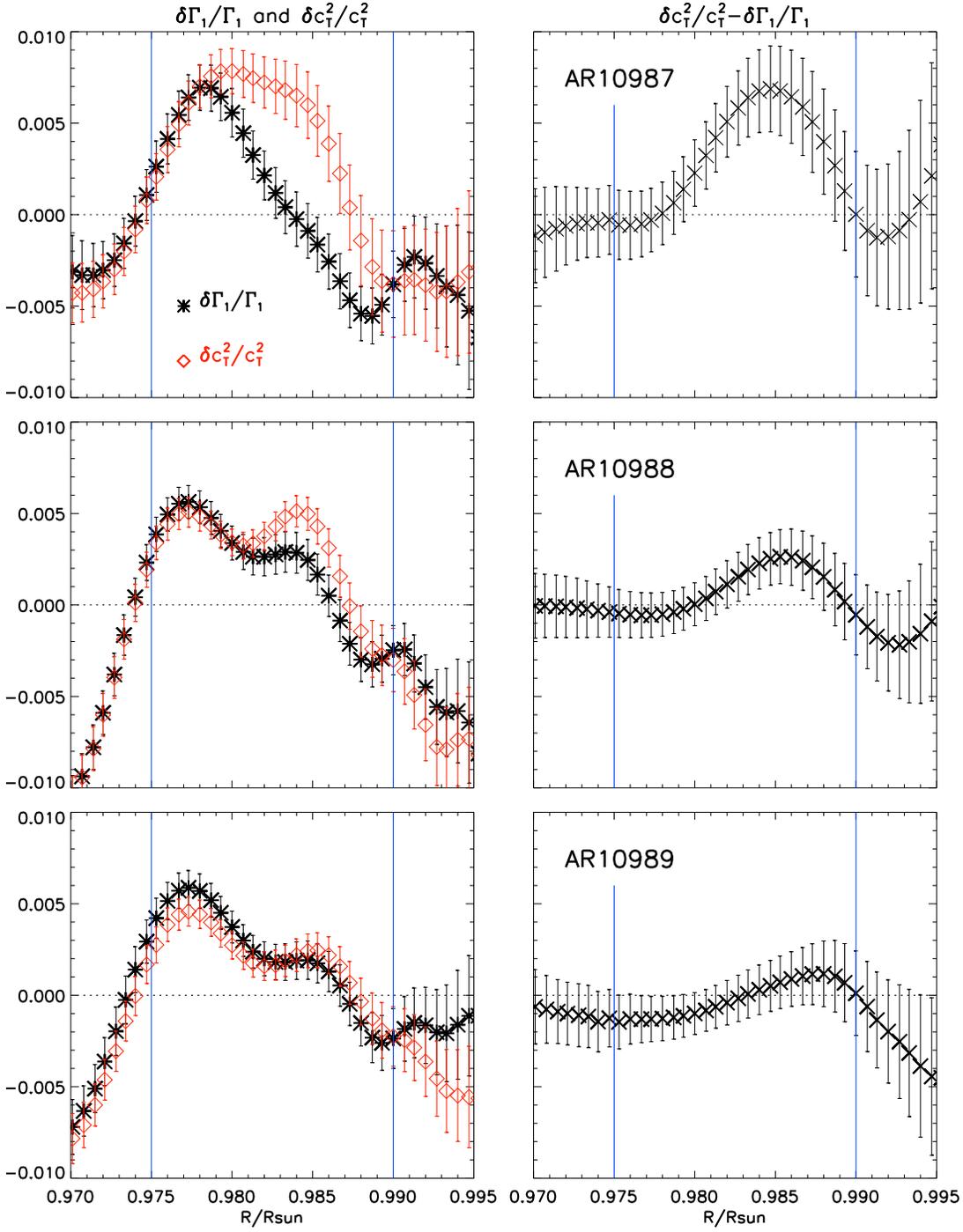}
\caption{
Same as in Fig.~\ref{fig:dg1dc2a} except that
the regions are AR10987--10989.
}
\label{fig:dg1dc2b}
\end{figure}

\begin{figure}
\includegraphics[width=15cm]{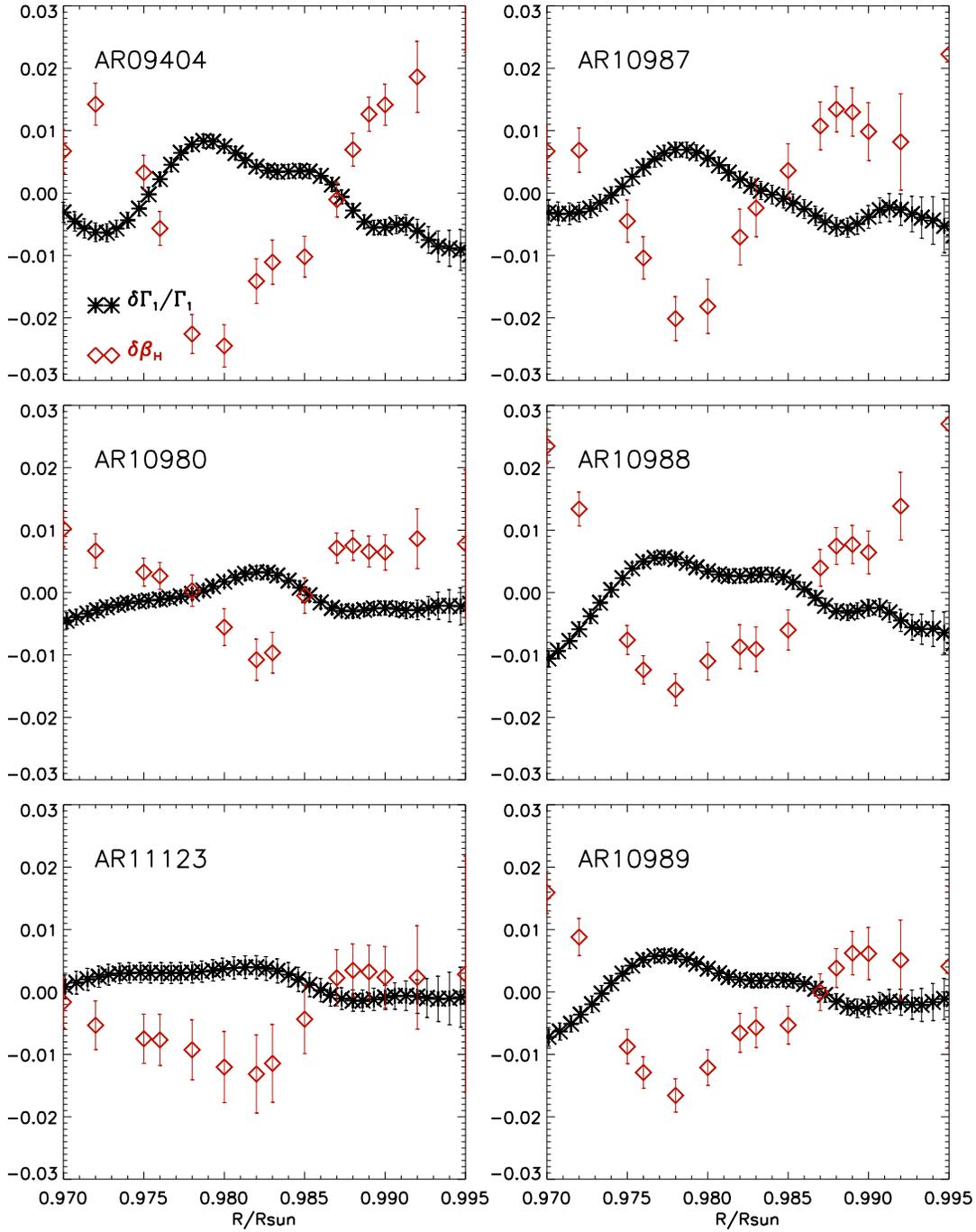}
\caption{The inferred subsurface 
$\delta \beta_{\rm H} (\equiv \delta(P_{\rm mag}/P_{\rm gas}))$ (red diamonds)
overplotted with 
the inversion results of $\delta \Gamma_1/\Gamma_1$ (black stars).
The corresponding active regions are as indicated.
}
\label{fig:beta}
\end{figure}

The inversion results of the six ARs are shown in 
Fig.~\ref{fig:dg1dc2a} (09404, 10980, 11123) and
Fig.~\ref{fig:dg1dc2b} (10987--10989).
The profiles of $\delta c_T^2/c_T^2-\delta \Gamma_1/\Gamma_1$
are displayed in the RHS column,
and $\delta c_T^2/c_T^2$ and $\delta \Gamma_1/\Gamma_1$
are plotted in the LHS column.
The ARs in both figures
are arranged from the {\em least} CME productive (top)
to the {\em most} CME productive regions (bottom).
The figures show that $\delta c_T^2/c_T^2-\delta \Gamma_1/\Gamma_1$
of most ARs reverses its sign first 
at a shallow layer around $0.985-0.99 R_\odot$
and again at a deeper point,
resulting in 
a positive bump below  $0.99 R_\odot$.
The positive $\delta c_T^2/c_T^2-\delta \Gamma_1/\Gamma_1$
means that the temperature at this layer is higher in an AR than 
in its reference QS region.
This feature has been interpreted as 
due to the accumulation of plasma that has been prohibited to
move into the magnetic field of the AR.
In the following discussion,
we will focus on the features in the depths within which
the plasma accumulation occurs and the inversion is most reliable;
that is, between $0.975 R_\odot$ and $0.99 R\odot$.
This region of interest is marked by the two vertical lines in
Figs.~\ref{fig:dg1dc2a} to \ref{fig:beta}.

We first inferred the subsurface temperature structure by
inspecting the profiles of
$\delta c_T^2/c_T^2 - \delta \Gamma_1/\Gamma_1$.
We found that 
the magnitude of $\delta c_T^2/c_T^2 - \delta \Gamma_1/\Gamma_1$
within $0.975-0.99 R_\odot$
decreases from 
top to bottom panels in both Fig.~\ref{fig:dg1dc2a} and \ref{fig:dg1dc2b},
suggesting a trend of smaller subsurface temperature difference
for more CME/flare productive AR.
This trend can be equivalently interpreted as
less plasma accumulation below more CME/flare productive ARs.
In the paper by \citet{Basuetal2004ApJ},
who studied
twelve selected ARs with different MAIs,
it was reported that the maximum magnitude of
$\delta c_T^2/c_T^2 - \delta \Gamma_1/\Gamma_1$,
within $0.975R_\odot-0.99R_\odot$, 
generally increases with increasing MAI.
However, two regions that had the highest MAIs in their study 
showed unexpectedly low value.
It was found that
these two regions, AR9026 and AR9393,
had the highest flare index
among all the ARs in their study.
Flare index
is a quantity to quantify the daily flare activity over 24 hours
per day \citep{FI_Antalova1996,AO1998SoPh,Kleczek1952},
Hence,
their results demonstrated another two flare productive ARs that
had unexpectedly small $|\delta c_T^2/c_T^2 - \delta \Gamma_1/\Gamma_1|$.
Based on these results,
we suggest that
higher flare/CME production
and
smaller subsurface ($0.975-0.99 R_\odot$) temperature differences
between the active regions and reference quiet Sun regions
may be correlated.
However, a statistical study is needed to verify this suggestion.

Next,
we use $\delta \Gamma_1/\Gamma_1$ profiles 
(black stars in the LHS panels)
to infer the subsurface magnetic structure.
Comparing $\delta \Gamma_1/\Gamma_1$ of different ARs 
between $0.975-0.99 R_\odot$,
we can see that the profiles of AR10988 and AR10989 resemble each other but
differ from the profiles of the other five ARs.
Since the resemblance
is only seen in two closely-located and CME-producing regions
but not among all eruptive regions,
we can deduce that it
may indicate a subsurface structural connection between the two ARs.
The inferred subsurface magnetic-field effect
(represented by $\delta \beta_H \equiv \delta(P_{\rm mag}/P_{\rm gas})$)
is plotted
in Fig.~\ref{fig:beta}.
Most part of a $\delta \beta_H$ profile resembles a reversed and magnified
$\delta \Gamma_1/\Gamma_1$. 
We caution that
this may be
due to the simplification of the magnetic field in 
the solar model on which the derivation is based.
The magnitude of the inferred $\delta \beta_H$ is likely to be 
an under-estimate of the actual value
\citep{lbl2009SoPh},
and the actual profile of $\delta \beta_H$ may contain complexities
that are not resolvable by our method.
A solar model implemented with more realistic magnetic fields
is required before better inference can be made.

\section{Summary} \label{sec:sum}
The objectives of this study were to
investigate the driving mechanisms of CMEs and
to search for possible subsurface structural signatures 
that may be correlated with the productivity.
Two large CMEs and six active regions,
including the source regions of the selected CMEs,
were investigated.

To determine the driving mechanisms,
we
tested various proposed models using observed CME data and 
associated SXR/HXR emission data.
Our analysis showed that
the EF model produced good fits to the kinematics of both CMEs.
The predicted foot-point separations at the bottom of corona were
consistent, within 20\%, with the sizes of the two source regions.
The predicted poloidal magnetic flux increasing rate ($d\Phi_p/dt$)
coincided with the SXR light curve associated with
CME03 (2008-03-25 CME),
but peaked slightly earlier than the SXR light curve of CME04 (2008-04-05).
This small discrepancy in time is consistent with the fact 
that the eruption of CME04
occurred slightly behind the limb from the viewpoint of GOES 
so that the SXR emissions were initially blocked by the solar limb.
However, some discrepancies between the EF model and CME04 were also found.
The model acceleration peaked higher and later than 
the observation-derived value.
The width of $d\Phi_p/dt$ is also greater than that of SXR emission profile.
These two inconsistencies imply that some assumptions related to
the acceleration and SXR emission mechanisms in the model 
are inappropriate for this event.

For the investigation of the subsurface structure,
we first implemented 
a local helioseismic inversion method to obtain
$\delta c_T^2/c_T^2$ and $\delta \Gamma_1/\Gamma_1$ between
active regions and their respective quiet-sun reference regions.
The subsurface temperature difference of each AR-QS pair was subsequently
deduced from
$\delta c_T^2/c_T^2 - \delta \Gamma_1/\Gamma_1$,
and
the difference of the subsurface magnetic structure
was inferred from 
$\delta \Gamma_1/\Gamma_1$. 
We found that
the maximum magnitude of the subsurface temperature difference
decreases with the CME/flare productivity of an AR.
That is,
the temperature difference is 
larger for less CME/flare productive regions and
smaller for more CME/flare productive regions.
An earlier study by \citet{Basuetal2004ApJ} also reported 
unexpectedly small temperature differences in the active regions
with highest flare indices among all their studied regions.
Based on the aforementioned results,
we suggest that there is
a correlation between the productivity of CME/flare
and the subsurface temperature structure.
We point out that
this correlation seems independent of
whether being CME/flare productive is a consequence or a cause
of the processes that reduce temperature difference.
The inferred $\delta \beta_H$ 
indicated
that the CME03 and CME04 source regions have similar subsurface structures.
The similarity may be an indication of subsurface structural connection
between the two regions.

The helioseismic inversion technique used in our study reveals
a static picture of the subsurface structures averaged over several days
(4 -- 6.5 days in our case).
The technique thus cannot resolve dynamical effects 
that occur on a time scale shorter than several days.
Since the time scale for observed CME eruptions is only tens of minutes
and initial/pre-eruption disturbances may only become detectable
hours prior to the eruptions,
we cannot conclude from our analysis
the exact subsurface effects that are directly correlated 
with the production of CMEs.
For instance,
while 
the overall productivity of CMEs may be directly related to
the non-dynamical structural difference uncovered in our analysis,
it is equally possible that 
the CMEs are more effectively triggered by certain dynamical effects
that are associated with the revealed non-dynamical structure.
Nevertheless,
our results indicate a potential connection between the coronal eruption
and the subsurface structure.
As we discussed in Sec.~\ref{sec:cmemdl},
the mathematical formulations (or numerical simulation results) of
all three selected CME models
show
that CME eruption is driven by an increase of the poloidal magnetic flux of
the erupting flux rope.
The difference is that CA and BO models explicitly propose that
the increase has to result from coronal reconnection,
while EF model does not specify the source of the increase.
It could result from either an existing coronal field via reconnection or
a sub-photospheric source \citep{CK2010ApJ}.
Another possibility is that 
different non-dynamical subsurface structures may lead to
different coronal magnetic field structures that may either facilitate
or inhibit coronal eruptions.
Currently, none of these scenarios has been observationally tested or testable.
More advanced observation and analysis techniques are required to
resolve this issue.

\section*{Acknowledgments}

CHL wish to thank Sarbani Basu and Angelos Vourlidas 
for helpful inputs and suggestion.
This work is funded by
the NSC of ROC under grant NSC99-2112-M-008-019-MY3
and the MOE grant ``Aim for the Top University'' to 
the National Central University.
CSB is supported by a NASA Earth and Space Sciences fellowship NNX08AY41H.
JC is supported by the Naval Research Laboratory Base Program.
The SECCHI data are produced by an international consortium of 
the NRL, LMSAL and 
NASA GSFC (USA), RAL and 
Univ. Bham (UK), MPS (Germany), CSL (Belgium), IOTA and 
IAS (France).
The data for the helioseismic inversion are from
the Solar Oscillations Investigation/Michelson Doppler Imager (SOI/MDI) on board
the Solar and Heliospheric Observatory (SoHO)
and from the Helioseismic and Magnetic Imager (HMI) on board
the Solar Dynamics Observatory.
SoHO is a project of international cooperation between ESA and NASA.
MDI is supported by NASA grant NNX09AI90G to Stanford University.

\begin{table}
\begin{tabular}{l|ccccc}
AR & Tracking time (UT)  & Carrington rotation & latitude & longitude & 
$\Delta$MAI\\
\hline
09404 &  2001.03.29\_21:06 -- 04.04\_13:38 & 1974 & 4~S  & 102 &  53.8~G\\
 QS   &                                    & 1974 & 4~S  & 152 & \\ \hline 
10980 &  2008.01.04\_12:37 -- 01.10\_05:09& 2065 & 7~S  & 237 &  47.5~G\\
 QS   &                                  & 2065 & 7~S  & 267 & \\ \hline 
10987 &  2008.03.24\_18:27 -- 03.30\_21:59 & 2068 & 7~S  & 260 &  76.8~G\\
 QS   &                                  & 2068 & 7~S  & 305 & \\ \hline 
10988 &  2008.03.26\_13:18 -- 04.02\_00:50 & 2068 & 7~S  & 237 & 79.3~G\\
 QS   &                                  & 2068 & 7~S  & 305 & \\ \hline
10989 &  2008.03.28\_04:32 -- 04.03\_17:04 & 2068 & 11~S & 205 & 40.4~G \\
 QS   &                                  & 2068 & 11~S & 195 &  \\ \hline
11123 &  2010.11.10\_04:21 -- 11.14\_10:44 & 2103 & 22~S & 190 & 19.7~G \\
 QS   &                                  & 2103 & 22~S & 300 &  \\ \hline
\end{tabular}
\caption{Tracking times, coordinates and 
$\Delta$MAIs (Magnetic Activity Index) of the AR-QS pairs.
$\Delta$MAI is the MAI difference between an AR and its QS reference.}
\label{tab:tab_coord}
\end{table}

\clearpage

\end{document}